\documentclass[preprint,3p,12pt]{elsarticle}
\usepackage{mathrsfs}
\usepackage{amsmath}
\usepackage{stmaryrd}
\usepackage{bbding}
\usepackage{dcolumn}
\usepackage{graphicx}	
\usepackage{amsfonts}
\usepackage{amssymb}
\usepackage{psfrag}
\usepackage{wrapfig}
\usepackage{subfigure}
\usepackage{makeidx}
\usepackage{bm}
\usepackage{epsf}
\usepackage{epsfig}
\usepackage{setspace}
\usepackage{graphicx}
\usepackage{epstopdf}
\usepackage{psfrag}
\usepackage{subfigure}
\usepackage{color}
\usepackage{comment}
\usepackage{caption}
\renewcommand{\vec}[1]{{\boldsymbol{#1}}}

\begin{document}

\title{Wave-Particle Turbulence Simulation of Spatially Developing Round Jets: Turbulent Flow Modeling and Method Validation}

\author[HKUST1]{Xiaojian Yang}
\ead{xyangbm@connect.ust.hk}

\author[HKUST1,HKUST2,HKUST3]{Kun Xu\corref{cor1}}
\ead{makxu@ust.hk}

\address[HKUST1]{Department of Mathematics, Hong Kong University of Science and Technology, Clear Water Bay, Kowloon, Hong Kong}
\address[HKUST2]{Department of Mechanical and Aerospace Engineering, Hong Kong University of Science and Technology, Clear Water Bay, Kowloon, Hong Kong}
\address[HKUST3]{Shenzhen Research Institute, Hong Kong University of Science and Technology, Shenzhen, China}
\cortext[cor1]{Corresponding author}

\begin{abstract}

Spatially developing round jet flows are fundamental to numerous engineering applications. This letter applies the wave-particle turbulence simulation (WPTS) method, a recently developed multiscale approach, to simulate a spatially developing circular jet at Reynolds number $5000$ — a canonical configuration for validating turbulence models. The study aims to further establish the effectiveness and accuracy of WPTS for shear-driven turbulent flows.
WPTS employs a multiscale framework that couples wave and particle components, where the wave component captures cell-resolved flow structures while the particle component models sub-grid flow information through non-equilibrium transport. Using a computational grid containing only $2\%$ of the cells required for direct numerical simulation (DNS), WPTS successfully predicts both qualitative flow features and quantitative turbulence statistics, including: (1) the characteristic linear decay of centerline velocity, (2) radial profiles of mean velocity, and (3) anisotropic Reynolds stress components. The results demonstrate excellent agreement with DNS data and experimental measurements. These findings establish WPTS as an effective computational tool for turbulence simulation, capable of maintaining high fidelity in capturing essential turbulence characteristics while utilizing significantly coarser grids than traditional methods. The successful application to turbulent jet flow demonstrates the method's potential for extension to more complex engineering flow configurations, offering a computationally efficient alternative for industrial applications.

\end{abstract}

\begin{keyword}
wave-particle turbulence simulation, non-equilibrium transport, jet flow
\end{keyword}

\maketitle

\section{Introduction}

Jet flows are fundamental to numerous engineering applications, particularly in aerospace propulsion systems, where their dynamics critically influence performance and efficiency. The study of jet flows holds both academic significance and practical importance for engineering design. Given the high Reynolds numbers characteristic of such flows, turbulence plays an essential role, making accurate prediction both challenging and crucial.

Experimental and numerical approaches have been extensively employed to investigate jet flow dynamics \cite{Tur-case-jet-exp-hussein1994velocity, Tur-case-jet-exp-panchapakesan1993turbulence}. Among numerical methodologies, direct numerical simulation (DNS), large-eddy simulation (LES), and Reynolds-averaged Navier-Stokes (RANS) modeling constitute the predominant computational techniques.
DNS resolves all turbulent scales through exceptionally refined computational grids and high-order numerical schemes, solving the Navier-Stokes equations without turbulence modeling assumptions. This approach yields highly accurate predictions of mean flow characteristics, second-order turbulence statistics, and higher-order quantities including triple velocity correlations \cite{Tur-case-jet-DNS-3rd-statistics-taub2013direct}. DNS investigations have demonstrated remarkable agreement with experimental measurements, establishing its reliability for fundamental turbulence research. Representative applications include: validation of self-similarity in fully developed turbulence \cite{Tur-case-jet-DNS-sharan2021investigation, Tur-case-jet-DNS-uPdfAna-nguyen2024analysis}, examination of high-pressure inflow effects \cite{Tur-case-jet-DNS-sharan2021investigation}, analysis of fluid residence time \cite{Tur-case-jet-DNS-7290-shin2017self}, and characterization of velocity probability density functions \cite{Tur-case-jet-DNS-uPdfAna-nguyen2024analysis}. Furthermore, DNS provides comprehensive flow field data essential for turbulence model development, exemplified by studies of third-order statistics in round jets at $Re_j = 2000$ \cite{Tur-case-jet-DNS-3rd-statistics-taub2013direct}. However, computational costs scale prohibitively with Reynolds number, limiting current DNS applications to moderate values such as $Re_j = 5000$ \cite{Tur-case-jet-DNS-sharan2021investigation} and $7290$ \cite{Tur-case-jet-DNS-7290-shin2017self}. Extension to engineering-scale flows remains computationally prohibitive.

To address computational constraints, LES and RANS approaches have been developed. LES solves filtered Navier-Stokes equations, resolving large-scale structures while modeling small-scale turbulence through subgrid-scale (SGS) models. This approach enables simulations at higher Reynolds numbers, reaching $Re_j = 1.71 \times 10^6$ \cite{Tur-case-jet-les-1e6-khosronejad2017experimental, Tur-case-jet-les-rediff-bogey2006large}, though accuracy remains dependent on SGS modeling. LES has been successfully applied to investigate noise generation \cite{Tur-case-jet-les-noise-wan2013large, Tur-case-jet-les-6e4-noise-bogey2003noise}, space-time velocity correlations \cite{Tur-case-jet-les-correla-zhang2019space}, and passive scalar intermittency \cite{Tur-case-jet-les-intermittency-gilliland2012external}. RANS, which fully models turbulence through eddy viscosity approximations, remains the industrial standard due to computational efficiency. While RANS effectively predicts mean flows, it struggles with unsteady flow features \cite{Tur-case-jet-rans-moddiff-sharma2023numerical}. The Reynolds stress model, which directly evolves Reynolds stresses, shows promise for strongly anisotropic flows like round jets \cite{Tur-case-jet-rans-reystress-turutoglu2024calibration}. Additional methods include detached eddy simulation \cite{Tur-case-jet-des-zhou2025enhanced}, transported PBE-PDF approaches \cite{Tur-case-jet-PBEPDF-di2011modeling}, and one-dimensional turbulence models \cite{Tur-case-jet-odt-sharma2022features}.

This paper applies the recently proposed wave-particle turbulence simulation (WPTS) method \cite{Tur-wpts-first-yang2025wave} to spatially developing jet flows. WPTS originates from the unified gas-kinetic wave-particle (UGKWP) framework \cite{WP-first-liu2020unified, UGKS-book-framework-xu2021cambridge, WP-gasparticle-polydis-yang2024unified, WP-radiation-implicit-liu2023implicit, WP-plasma-pip-pu2025unified, WP-ada-wei2023adaptive}, based on direct modeling concepts from the unified gas-kinetic scheme (UGKS) \cite{UGKS-xu2010unified, UGKS-book-xu2014direct, UGKS-multigrid-zhu2017unified, UGKS-implicit-ada-long2024implicit}.
The UGKWP method's core strength lies in its ability to capture nonequilibrium transport in rarefied flow regimes where numerical cell resolution cannot adequately resolve the particle mean free path scale. This provides a multiscale framework that seamlessly captures flow physics from rarefied to continuum regimes without the inherent constraints of the direct simulation Monte Carlo (DSMC) method. This methodology naturally extends to turbulent flow simulation, where turbulent kinetic energy is represented through breaking down fluid elements or particles exhibiting nonequilibrium transport within computational cells that cannot fully resolve flow structures. In this framework, particle generation and annihilation processes correspond to transitions between turbulent and laminar flow states, providing an adaptive mechanism that responds to local flow conditions. The particle dynamics thus intrinsically model subgrid-scale turbulent energy production, transport, and dissipation, while the wave component captures resolved flow structures, creating a self-consistent multiscale representation of turbulent flows.

\begin{figure}[htbp]
	\centering
	\subfigure{
		\includegraphics[height=6.7cm]{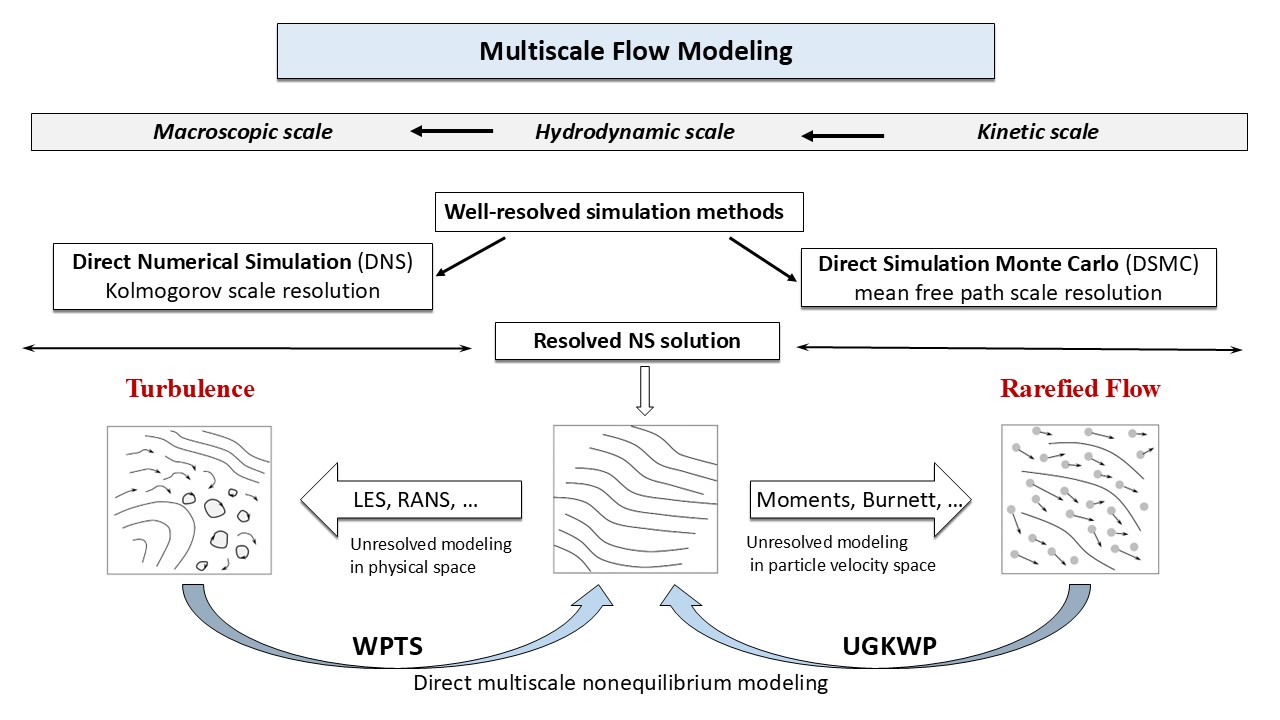}
	}
	\caption{The illustration of the correlation of rarefied flow and turbulence based on multi-scale modeling.}
	\label{Fig-picture}
\end{figure}

Figure \ref{Fig-picture} illustrates the multiscale modeling analogy between rarefied flows and turbulence. For rarefied flows, DSMC resolves mean free path scales, with NS solutions emerging through particle collisions. Analogously, DNS resolves Kolmogorov scales in turbulence, with coarse-grid NS solutions representing high-dissipation limits. Both UGKWP and WPTS employ wave-particle frameworks for non-equilibrium transport: away from fundamental scales (mean free path or Kolmogorov scale), unresolved physics exhibits cell-size-dependent multiscale characteristics associated with fluid element transport.

Unlike UGKWP for rarefied flows \cite{WP-first-liu2020unified, WP-second-zhu-unstructured-mesh-zhu2019unified, WP-sample-xu2021modeling}, where particles represent molecules with fluctuations corresponding to internal energy, WPTS treats particles as discrete fluid elements whose variance represents subgrid turbulent kinetic energy. The method introduces a hybrid framework decomposing turbulent flow into grid-resolved wave structures and stochastically modeled particle structures. The wave-particle decomposition dynamically adapts to local resolution and turbulence intensity: in well-resolved regions, WPTS reduces to Navier-Stokes equations, while in under-resolved turbulent regions, particles capture unresolved fluid information through free transport modeling. Our previous validation on temporally evolving mixing layers \cite{Tur-wpts-first-yang2025wave} demonstrated WPTS accuracy on coarse grids. Here, we extend validation to spatially developing jets—a more complex and practically relevant configuration.

The paper is organized as follows: Section 2 outlines WPTS methodology; Section 3 presents numerical setup, results, and discussion for jet simulations; and Section 4 provides conclusions.

\section{Wave Particle Turbulent Simulation (WPTS) Method}
The WPTS is a multiscale wave-particle coupled evolution framework developed based on the finite volume method (FVM). Importantly, this framework decomposes the flow field into two components: (1) wave-like structures, which represent large-scale flow features resolvable at the grid level, and (2) discrete fluid particles, which directly model the evolution of subgrid-scale flow dynamics mainly based on the non-equilibrium transport. Specifically, the wave and particle components are governed by Eulerian and Lagrangian descriptions, respectively, and their coupled evolution is formulated through a relaxation-type kinetic equation
\begin{gather}\label{bgk}
\frac{\partial f}{\partial t}
+ \nabla_x \cdot \left(\vec{u}f\right)
= \frac{g-f}{\tau},
\end{gather}
where $f$ is the probability distribution function (PDF) of molecules, $\vec{u}$ is the velocity, $\tau$ is the collision or relaxation time, and $g$ is the equilibrium state with
\begin{gather}\label{geq}
g = \rho \left(\frac{\lambda}{\pi} \right)^{\frac{K+3}{2}} e^{-\lambda \left[\left(\vec{u} - \vec{U}\right)^2 + \vec{\xi}^2 \right]}.
\end{gather}
$K$ is the internal degree of freedom taken as 2 for the diatomic molecule gas.

In WPTS, laminar and turbulent are simulated in a unified way. Particularly, only wave exists in the evolution for the laminar region (or the turbulence flow with a fine-enough grid) while stochastic particles will automatically appear for turbulence flow with a coarse grid, whose size can not resolve the minimum turbulence flow structures.
With the above in mind, $\lambda = \frac{1}{2RT}$ where $T$ stands for the thermal temperature $T_{thermal}$ only for laminar, but it also includes the turbulence temperature $\Theta_t$ in the case of turbulence. The turbulence temperature $\Theta_t$ is determined by the TKE, $\rho E_t$, through $\rho E_t = \frac{3}{2}\rho \Theta_t$. Accordingly, the total energy $\rho E$ for laminar includes the kinetic energy $\frac{1}{2}\rho \vec{U}^2$ and thermal energy $\rho R T_{thermal}$, while for turbulence it also includes $\rho E_t$ as well.
It is worth noting that, take moments with $g$ by $\vec{\psi}$ and then the macroscopic conservative variables, $\vec{W}=\left(\rho, \rho \vec{U}, \rho E\right)^T$, can be obtained, where $\vec{\psi}=(1,\vec{u},\displaystyle \frac{1}{2}\left(\vec{u}^2+\vec{\xi}^2\right))^T$.
The conservation law is satisfied based on the compatibility condition
\begin{gather*}
\int \vec{\psi}\left(g - f\right)\text{d}\vec{\Xi} = \vec{0},
\end{gather*}
where $\text{d}\vec{\Xi}=\text{d}u\text{d}v\text{d}w\text{d}\xi_1...\text{d}\xi_{K}$.

The incorporation of TKE into the kinetic equation provides a potential foundation for a unified description of laminar and turbulent flows. However, in practical computations, obtaining exact solutions for turbulent flows under coarse-grid resolution (which means the small-scale flow structures are inherently existing) remains theoretically intractable. Consequently, in WPTS we adopt the idea of direct modeling, whose the core lies in constructing a multiscale flux based on the non-equilibrium transport. Specifically, this flux is determined by both wave and particle components, as detailed below.

Starting from the relaxation-type kinetic equation, the integral solution of the PDF can be obtained,
\begin{equation}\label{bgk-integrasol}
f(\vec{x},t,\vec{u})=\frac{1}{\tau}\int_0^t g(\vec{x}',t',\vec{u} )e^{-(t-t')/\tau}\text{d}t'\\
+e^{-t/\tau}f_0(\vec{x}-\vec{u}t, \vec{u}),
\end{equation}
where $\vec{x}'=\vec{x}+\vec{u}(t'-t)$ is the trajectory of particles, $f_0$ is the initial gas distribution function at time $t=0$.
It indicates that the evolution of PDF consists of two distinct components: (1) the cumulative equilibrium effects along the integration path, and (2) the free-streaming of the initial distribution function along the characteristic line.

Our prior experience on the construction of multiscale flux for non-equilibrium flow, such as rarefied flow, radiation transport, etc., demonstrates that the equilibrium component represents the macroscopic fluid behaviors. It also determines the solution of the multiscale system in the equilibrium limit.
Therefore, in WPTS currently it is described by the Navier-Stokes (NS) equations, which ensures the automatic recovery of the NS solution for laminar flow.
So, the computation of flux from the equilibrium part, denoted as $\vec{F}^{eq}$, follows the gas-kinetic scheme (GKS) methodology. Specifically, the distribution function of the equilibrium state at the interface can be constructed based on the Taylor expansion in space and time. In summary, the calculation of this part has three steps as the standard procedure in FVM:
\begin{enumerate}
	\item Reconstruction: interpolating cell-averaged conserved variables $\vec{W}$ to interfaces. In this paper, the fifth-order WENO-AO reconstruction is employed.
	\item Evolution: computing fluxes as $\vec{F}^{eq}$ ,
	\begin{gather}\label{eqFluxeq}
	\vec{F}^{eq}_{ij}
	=\frac{1}{\Delta t} \int_{0}^{\Delta t} \int \vec{u}\cdot\vec{n}_{ij} f_{ij}^{eq}(\vec{x},t,\vec{u})\vec{\psi}\text{d}\vec{u}\text{d}t,
	\end{gather}
	where $f^{eq}$ is the time-dependent PDF of the equilibrium state at the interface,
	\begin{gather}\label{eqfeq}
	f^{eq}(\vec{x},t,\vec{u}) \overset{def}{=} \frac{1}{\tau}\int_0^t g(\vec{x}',t',\vec{u})e^{-(t-t')/\tau}\text{d}t' \nonumber\\
	= c_1 g_0\left(\vec{x},\vec{u}\right)
	+ c_2 \overline{\vec{a}} \cdot \vec{u} g_0\left(\vec{x},\vec{u}\right)
	+ c_3 A g_0\left(\vec{x},\vec{u}\right),
	\end{gather}
	with coefficients,
	\begin{align}\label{coefeq}
	c_1 &= 1-e^{-t/\tau}, \notag \\
	c_2 &= \left(t+\tau\right)e^{-t/\tau}-\tau, \\
	c_3 &= t-\tau+\tau e^{-t/\tau} \notag .
	\end{align}	
	\item Projection: updating the cell-averaged conserved variables $\vec{W}$. Since both the equilibrium and free-transport flux are required in the projection, the complete update expression for the conserved variables ($\vec{W}$) will be derived systematically following the detailed presentation of all flux contributions.
\end{enumerate}

The free-streaming component $e^{-t/\tau}f_0(\vec{x}-\vec{u}t, \vec{u})$ in Eq.(\ref{bgk-integrasol}) constitutes the fundamental origin of non-equilibrium transport phenomena. Within the WPTS framework, this component is simulated through stochastic particles that represent the discrete fluid elements. As discussed in \cite{Tur-wpts-first-yang2025wave}, we assume that in regions of high turbulence intensity, part of fluid will breakdown their connection in the evolution, and thus the Lagrangian particle is taken for modeling the turbulence flow within the un-resolved coarse grid (where the cell size significantly exceeds the characteristic scale of the smallest turbulent structures).

Specifically, the governing equations for these particles adopt a relaxation-type formulation, with the pressure gradient currently serving as the sole external driving force
\begin{gather}\label{dudtpar}
\frac{\text{d} \vec{u}\left(t\right)}{\text{d} t} = \frac{\vec{U} - \vec{u}}{\tau_n} + \vec{a},
\end{gather}
and $\vec{a} = \frac{\nabla p}{\rho}$. The collision time $\tau_n$ in Eq.\eqref{dudtpar} is taken as the sum of physical and turbulence collision time, namely, $\tau_n = \tau + \tau_t$, where $\tau = \mu/p$ and $\tau_t$ will be introduced later. One influence of the introduction of $\tau_n$ is it can adjust the weight of equilibrium and free transport in the evolution. Therefore, accordingly, the $\tau$ in the exponential in Eq.\eqref{coefeq} should be represented by $\tau_n$.
Besides, the authors emphasize that this baseline model can be systematically enhanced by incorporating additional physical factors, thereby progressively improving both the accuracy and generality of the WPTS.
Leveraging the UGKWP method, the coupled wave-particle approach is employed for the computation of this part. Particularly, only selected particles (those that move in whole time step) are treated via Lagrangian particle tracking, and the flux of remaining components is computed using the wave representation, denoted as $\vec{F}_{fr}^{wave}$ \cite{WP-first-liu2020unified, Tur-wpts-first-yang2025wave, UGKS-book-framework-xu2021cambridge}.
The calculation of $\vec{F}_{fr}^{wave}$ follows the same procedure with $\vec{F}_{eq}$ described above.

Importantly, the calculation of stochastic particles in WPTS can be summarized as:
\begin{enumerate}
	\item Sample particles from the wave,
	\begin{gather}
	\vec{W}^{hp}_i = e^{-\Delta t/\tau_n} \vec{W}^{h}_i,
	\end{gather}
	with $t_f = \Delta t$, and the velocity of the sampled particle is determined by
	\begin{gather}
	\vec{u}_p = \delta \vec{u}_p + \vec{U}
	\end{gather}
	where $\delta\vec{u}_p = \mathcal{D}_{N} \left[C_0\left(1-e^{\Delta t/\tau_n}\right)\rho E_t, \rho^h\right]$, depending on the fluid density by wave, $\rho^h$, and modeled production of TKE, $C_0\left(1-e^{\Delta t/\tau_n}\right)\rho E_t$. More details about the particle sampling can be found in \cite{Tur-wpts-first-yang2025wave}.
	
	Determine the $t_f$ for the surviving particles from the last step (if existing),
	\begin{gather}
	t_f = \text{min}\left[-\tau_n\text{ln}\left(\eta\right), \Delta t\right].
	\end{gather}
	\item Move particles by operator splitting, which indicates,
	\begin{gather}
	\vec{x}^* = \vec{x}^n + \vec{u}^n t_f,
	\end{gather}
	for free streaming, and
	\begin{align}
	\vec{u}^{n+1} &= \vec{u}^n + \vec{a} t_f, \\
	\vec{x}^{n+1} &= \vec{x}^* + \frac{1}{2}\vec{a} t_f^2,
	\end{align}
	for the acceleration, and meanwhile count the flux caused by the movement of particles, denoted as $\vec{w}_{i}^{fr,part}$,
	\begin{gather}
	\vec{w}_{i}^{fr,part} = \sum_{k\in P\left(\partial \Omega_{i}^{+}\right)} \vec{\phi}_k - \sum_{k\in P\left(\partial \Omega_{i}^{-}\right)} \vec{\phi}_k,
	\end{gather}
	where $P\left(\partial \Omega_{i}^{+}\right)$ is the particle set moving into the cell $i$ during one time step, $P\left(\partial \Omega_{i}^{-}\right)$ is the particle set moving out of the cell $i$ during one time step, $k$ is the particle index in the set, and $\vec{\phi}_k=\left[m_{k}, m_{k}\vec{u}_k, \frac{1}{2}m_{k}\vec{u}^2_k + m_k\frac{K+3}{2} \frac{1}{2\lambda_k}\right]^T$ is the mass, momentum and energy carried by particle $k$.
	\item Delete particles with $t_f < \Delta t$, and the carried conserved variables will merge into $\vec{W}$.
	
	Calculate the TKE characterized by the surviving particles, $\rho E_t$.
\end{enumerate}
Up to now, we have obtained all the necessary components of the flux, enabling the subsequent update of conserved quantities.
\begin{gather}\label{particle phase equ_updateW_ugkp}
\vec{W}_i^{n+1} = \vec{W}_i^n
- \frac{\Delta t}{\Omega_i} \sum_{S_{ij}\in \partial \Omega_i}\vec{F}^{eq}_{ij}S_{ij}
- \frac{\Delta t}{\Omega_i} \sum_{S_{ij}\in \partial \Omega_i}\vec{F}^{fr,wave}_{ij}S_{ij}
+ \frac{\vec{w}_{i}^{fr,part}}{\Omega_{i}}.
\end{gather}
It is worth noting that, the update of $\vec{W}_i^{p}$ inside each cell can be obtained by summing the contributions from all particles survived inside the cell, and further $\vec{W}^h$ can be obtained based on the conservation $\vec{W}^{h}_i = \vec{W}_i^{n+1} - \vec{W}^p_i$.
In WPTS, the TKE is directly evolved based on the particles' evolution, such as sampling, movement, deletion, etc. As a result, the governing equation of TKE in WPTS can not be written explicitly.

In WPTS, the turbulence collision time $\tau_{t}$ is crucial, since (1) it governs the relaxation timescale, directly influencing the particle's transport time, and (2) it also determines the percentage of particle component in wave-particle decomposition.
As a result, the model of $\tau_{t}$ is essential for the computational accuracy and efficiency across different flow regimes.
In this paper, the model proposed previously is employed here
\begin{align}\label{tautur}
e^{-\Delta t / \tau_t} = \left(1-\omega_p\right) e^{-\Delta t / \tau_{mac}} + \omega_p E_p,
\end{align}
which simultaneously incorporates the effects from both macroscale resolved component and stochastic particle component, and the $\tau_{mac}$ is evaluated based on the Smagorinsky model.
In this work, the coefficients in $\tau_t$ are $C_s^2 = 0.05$, $E_p = 0.8$, $k$ takes 0.98 when $\sqrt{\Theta_t} > 10^{-10}$ and zero otherwise. More details and discussion about $\tau_t$ can be found in \cite{Tur-wpts-first-yang2025wave}.

\section{Numerical simulation}	
In this section, the numerical simulation for round jet flow by the WPTS method and results will be introduced, including the computational setup, boundary conditions, the typical flow features, etc.

\subsection{Numerical setup}
As the treatments of WPTS in \cite{Tur-wpts-first-yang2025wave}, the fifth-order WENO-AO reconstruction is adopted for $\vec{W}$, by which the flux of equilibrium state $\vec{F}^{eq}$ is determined \cite{GKS-HLLC-compare-yang2022comparison, wenoao-gks-ji2019-performance-enhancement}; while the second-order reconstruction with van Leer limiter is employed for $\vec{W}^h$ by which the flux $\vec{F}^{fr,wave}$ can be obtained. For the temporal discretization, the wave component is evolved by the two-step fourth-order method, while the stochastic particle component is transported as described above in a whole time step $\Delta t$.
Furthermore, for the cell without any particle, the flux $\vec{F}$ of GKS will be automatically used \cite{GKS-2001}, indicating the WPTS will go back to the exact high-order GKS in these cells, such that the NS solution for laminar flow is obtained.
In this paper, the reference mass of one stochastic particle is $10^{-3}\Omega$. Besides, the CFL number is taken as 0.3, and $C_0$ is taken as 0.5, which are the same as our previous study \cite{Tur-wpts-first-yang2025wave}.

\subsection{Case setup}
For the study of round jet problems, the flow state is commonly characterized by two key dimensionless parameters: the jet Reynolds number $Re_j$ and the jet Mach number $Ma$. The Reynolds number is defined as:
\begin{gather*}
Re_j = \rho U_e D / \mu_e,
\end{gather*}
where $\rho$ is the density of jet flow, $U_e$ is the velocity at the core of jet exit, $D$ denotes the nozzle diameter, and $\mu_e$ represents the viscosity of the jet flow. In the simulation, the viscosity is calculated based on the power law, $\mu = \mu_e \left(\frac{T}{T_e}\right)^\omega$ with $\omega=0.667$. The Mach number is given by:
\begin{gather*}
Ma = U_e / c_e,
\end{gather*}
where $c_e$ is the speed of sound in the jet inflow. In the present study, a jet case with $Re_j = 5000$ and $Ma = 0.6$ is simulated by WPTS. Further details regarding this problem studied by DNS can be found in the reference \cite{Tur-case-jet-DNS-sharan2021investigation}.

\begin{figure}[htbp]
	\centering
	\subfigure{
		\includegraphics[height=5.0cm]{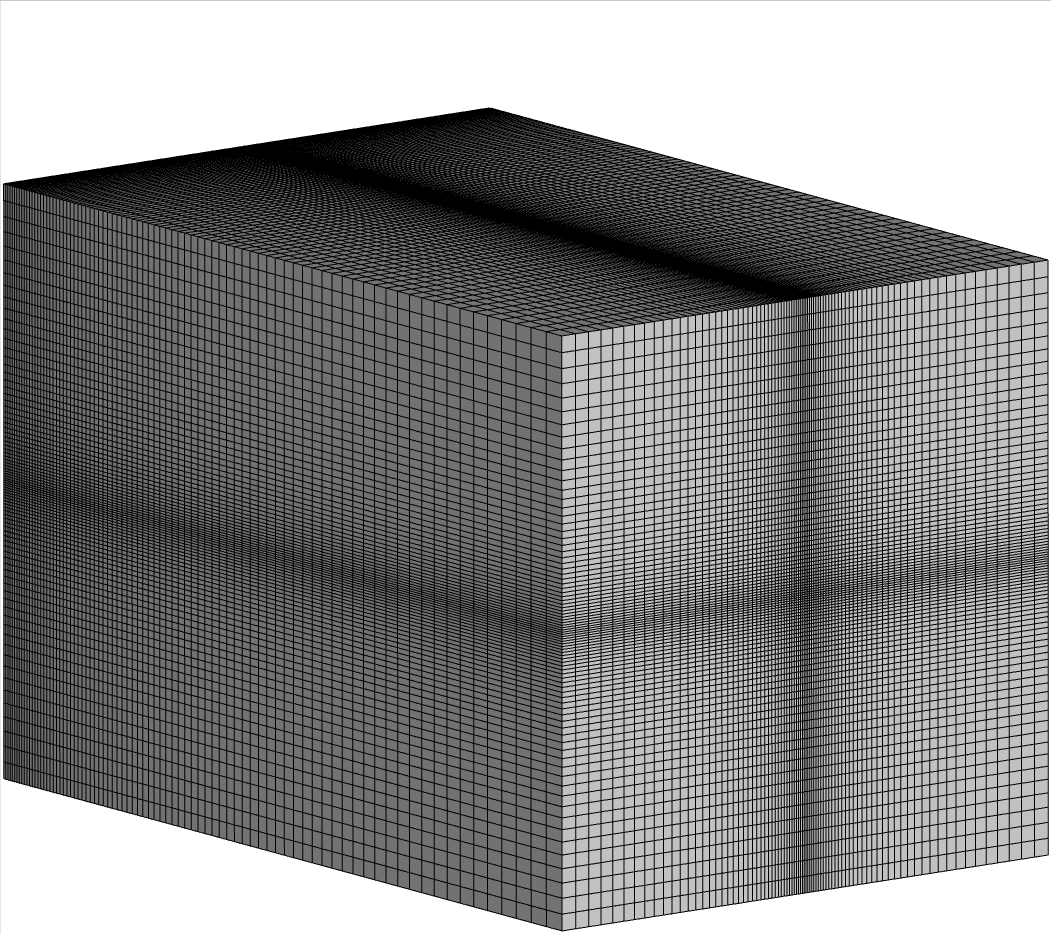}
	}
	\subfigure{
		\includegraphics[height=4.3cm]{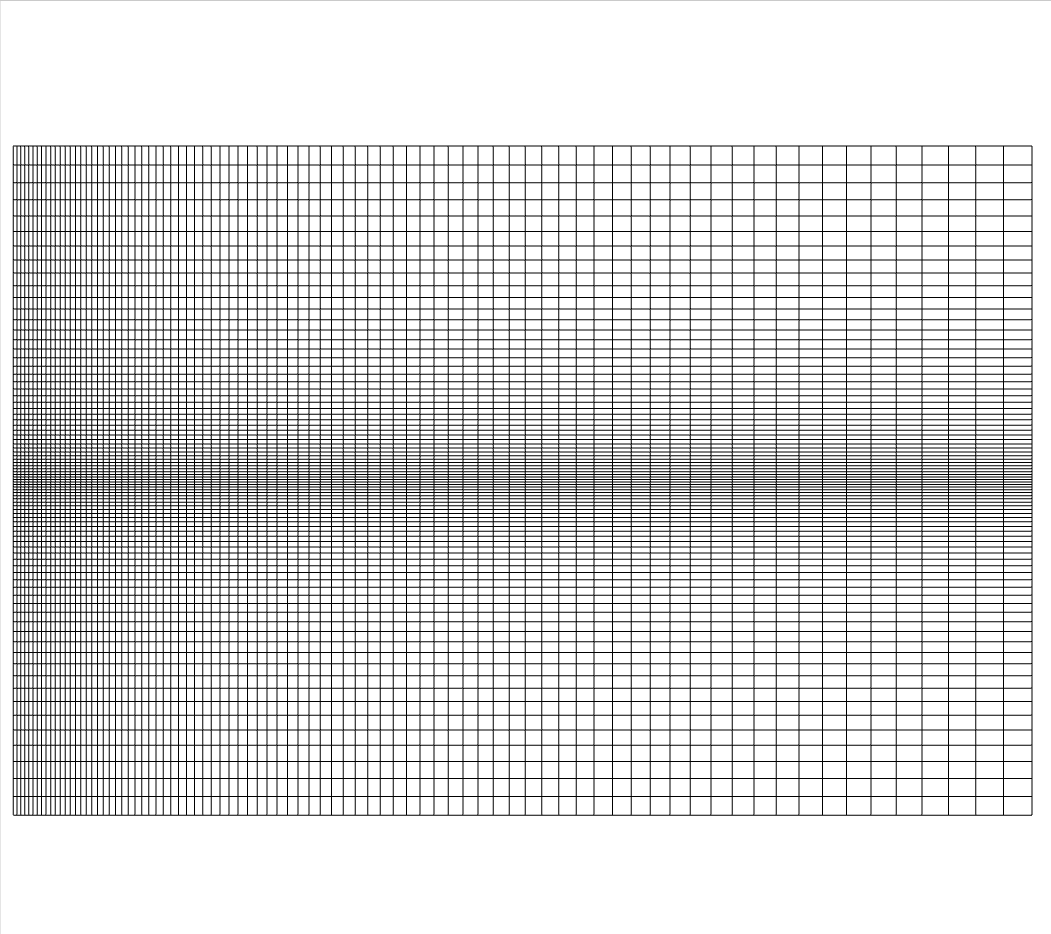}
	}
	\caption{The mesh employed for the round jet case: (left) three-dimensional view, (right) two-dimensional view at $xoy$ plane.}
	\label{Fig-mesh}
\end{figure}

The computational domain extends $45D$ and $30D×\times30D$ in the streamwise $\left(x\right)$ and transverse $\left(y, z\right)$ directions, respectively. A non-uniform Cartesian grid is employed, which is illustrated in Figure \ref{Fig-mesh}. The minimum grid spacings are $\Delta x_{min} = 0.15D$ in the streamwise direction and $\Delta y_{min} = \Delta z_{min} = 0.10D$ in the transverse directions. As shown in Figure \ref{Fig-mesh}, the cell with minimum volume is around the center of the jet exit. Besides, the grid is stretched exponentially along the streamwise and transverse directions with growth rates of 1.025 and 1.052, respectively. The total number of grid points is $84^3$, corresponding to approximately $2 \%$ that employed in DNS study \cite{Tur-case-jet-DNS-sharan2021investigation}.

The specification of inflow conditions for jet simulations has been extensively investigated in many studies. Typically, besides the mean velocity profile, the perturbations, such as white noise \cite{Tur-case-jet-inflow-landa2004development}, time-dependent fluctuations \cite{Tur-case-jet-inflow-varadharajan2017reynolds}, artificially generated velocity data \cite{Tur-case-jet-inflow-genedata-klein2003digital, Tur-case-jet-inflow-wang2010direct}, and specific excitation modes \cite{Tur-case-jet-inflow-gohil2015simulation, Tur-case-jet-inflow-gohil2019numerical}, are usually superimposed at the inlet to trigger turbulence and accelerate transition to a fully developed state. Alternatively, in some studies, the flow fields developed through the pipe are also employed as inflow conditions to better match realistic jet features \cite{Tur-case-jet-DNS-sharan2021investigation, Tur-case-jet-DNS-uPdfAna-nguyen2024analysis}.
Nevertheless, for round jets, both experimental and numerical studies have demonstrated the existence of self-similarity in the fully developed turbulent region, which serves as an important validation criterion in the current numerical study.

In the current work, the mean velocity profile at the inflow boundary is prescribed by a hyperbolic tangent function \cite{Tur-case-jet-DNS-sharan2021investigation}:
\begin{gather*}
U(r)=\frac{U_e}{2}\left[1-\text{tanh}(\frac{r-r_0}{2\theta_0})\right], ~~ V=0, ~~W=0,
\end{gather*}
where $r_0 = D/2$ is the jet radius, $\theta_0$ is the initial momentum thickness taken as $0.04r_0$, and $r$ is defined as $r = \sqrt{y^2+z^2}$.
Besides, for the excitation of turbulence, the dual-mode perturbations $u(r,t)$ are superimposed on the streamwise velocity $U$ within the jet core $r\leq r_0$,
\begin{gather}\label{eq-uInletmode}
u(r,t)=A_n U_e \text{sin}\left(2 \pi St_D \frac{U_e}{D} t\right)
+ A_h U_e \text{sin}\left(2 \pi St_H \frac{U_e}{D} t - \theta \right)\left(\frac{2r}{D}\right),
\end{gather}
where $\theta = \text{atan}\left(y,z\right)$ is the azimuthal direction, the amplitudes are $A_n=A_h=0.05$, the axisymmetric Strouhal number is $St_D = 0.5$, and the helical Strouhal number is $St_H = St_D / f$ with the frequency ratio $f = 2.40$, which are chosen based on the previous study \cite{Tur-case-jet-inflow-gohil2015simulation}.
Finally, the small perturbations which are uniformly distributed in $[-kU_e, kU_e]$ are further superimposed for three velocity components in the jet core $r\leq r_0$, and $k$ is taken as 0.004.

For the downstream boundary, a convective outflow condition is imposed, while non-reflecting boundary conditions are applied at the transverse boundaries.
In the simulation, $D$ is taken as a unit. Besides, the characteristic flow timescale is defined as $T_e = L/U_e$ where $L=45D$ is the streamwise extent of the computational domain.

\subsection{Results and discussion}

\begin{figure}[htbp]
	\centering
	\subfigure{
		\includegraphics[height=6.0cm]{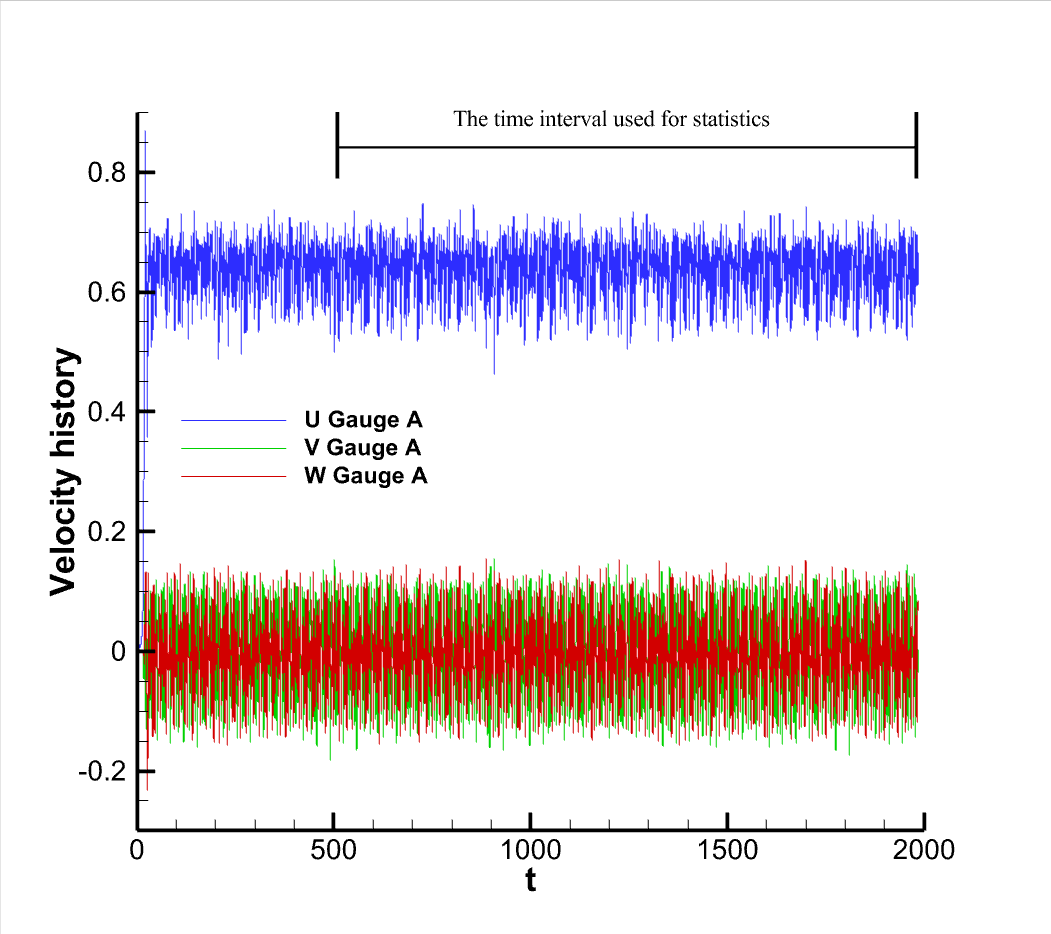}
	}
	\quad
	\subfigure{
		\includegraphics[height=6.0cm]{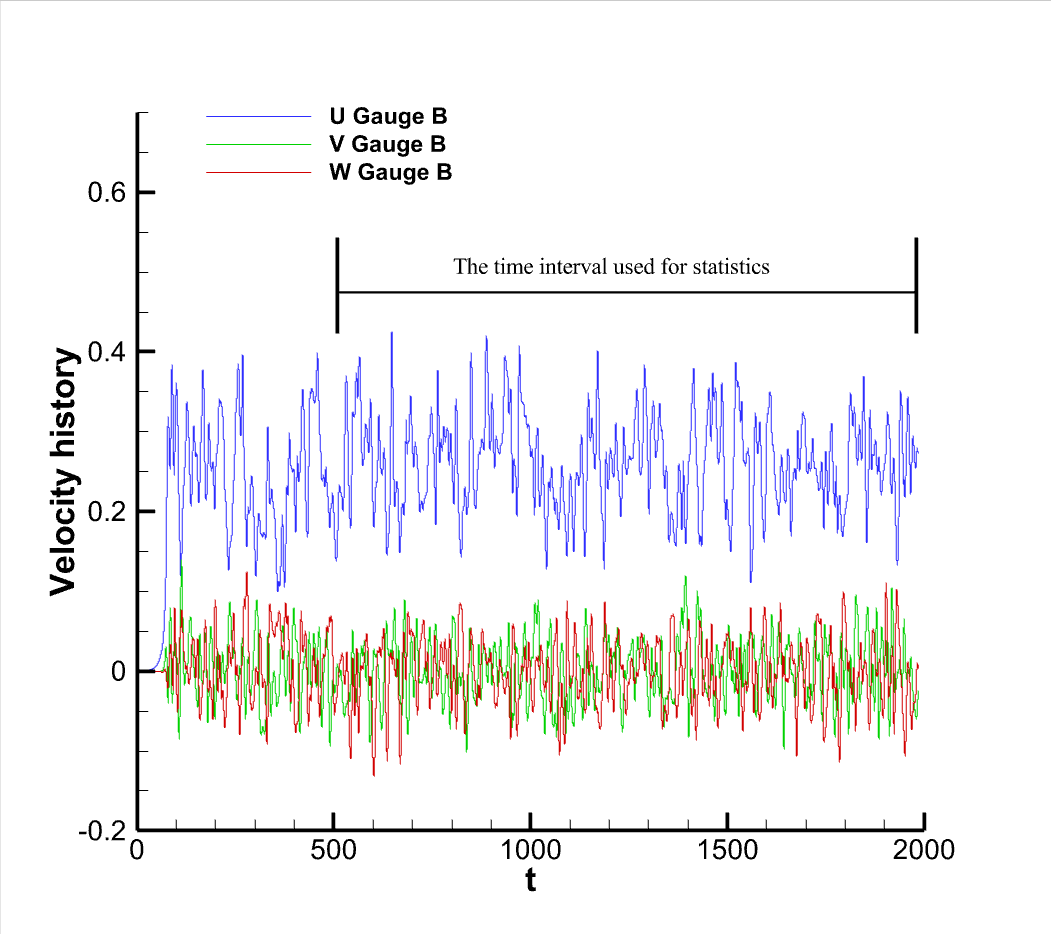}
	}
	\caption{The history of velocities at gauge points: (left) gauge point A at (4.8, 0, 0), (right) gauge point B at (16.5, 0, 0).}
	\label{Fig-Gauge}
\end{figure}

The numerical simulation spans a dimensionless time of $t=2000$, corresponding to approximately $31.5T_e$. In the simulation, two monitoring points were established at locations A (4.8, 0, 0) and B (16.5, 0, 0), and Figure \ref{Fig-Gauge} illustrates the temporal evolution of all three velocity components at these positions. Generally, the velocity fluctuations exhibit typical turbulent features, oscillating within a range around their mean values. The final $23T_e$ in the simulation, which is marked in Figure \ref{Fig-Gauge}, is utilized for the following statistical analysis of turbulent quantities, such as the averaged flow variables, the Reynolds stress components, etc.

\subsubsection{Self-similarity}
\begin{figure}[htbp]
	\centering
	\subfigure{
		\includegraphics[height=6.5cm]{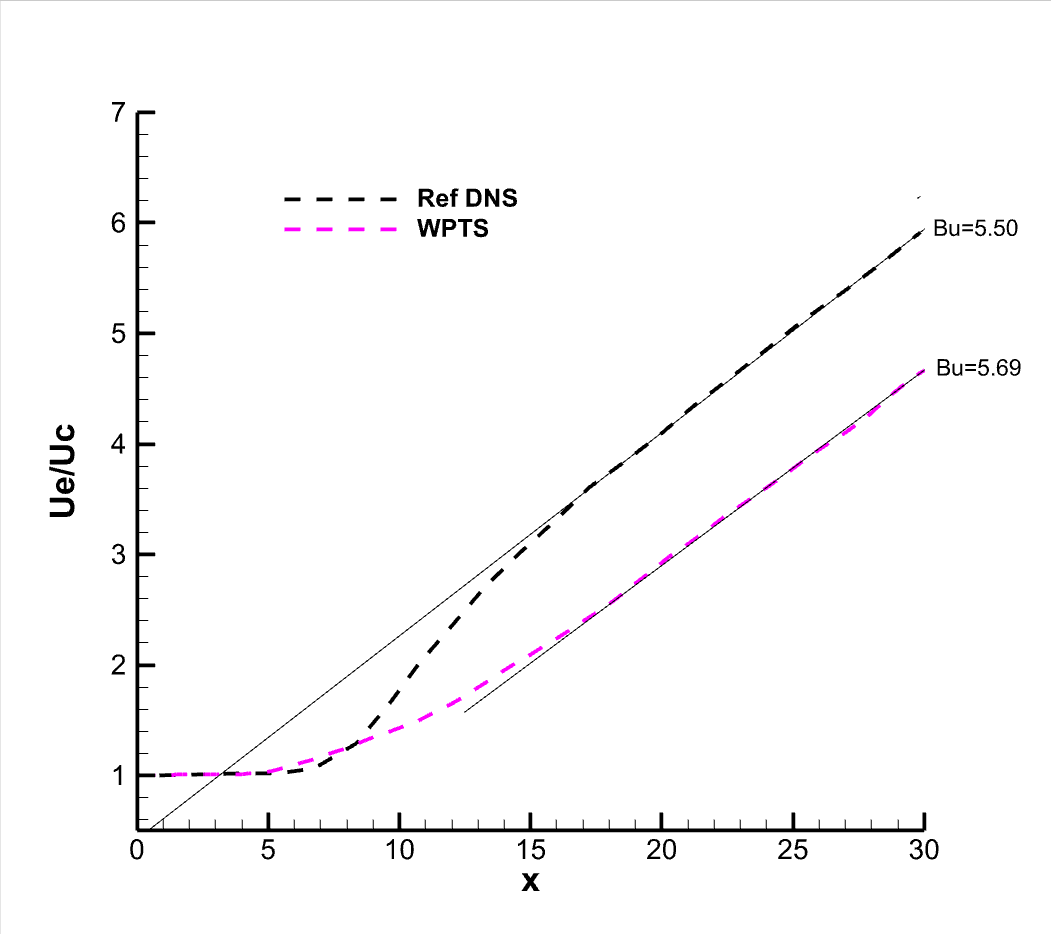}
	}
	\caption{The inverse of averaged centerline mean velocity, and the reference is from \cite{Tur-case-jet-DNS-sharan2021investigation}.}
	\label{Fig-Bu}
\end{figure}

As introduced above, for the round jet case, the self-similarity feature in the fully-developed turbulence region has been found in many previous studies, and it will be checked in this work.
Firstly, one key feature of the self-similarity is the linear decay of the mean streamwise velocity at the centerline, which can be expressed as:
\begin{equation}
\frac{U_e}{U_c(x)} = \frac{B_u D}{x - x_{0u}},
\end{equation}
where $U_c(x)$ denotes the centerline mean velocity at various streamwise positions, $B_u$ characterizes the decay rate, and $x_{0u}$ represents the virtual origin of the jet. For the fully developed turbulent state in jet flow, $B_u$ should remain constant. As demonstrated in Figure \ref{Fig-Bu}, the WPTS method accurately captures this linear decay feature. For the quantitative comparison, Table \ref{Tab-Bu} presents the value of $B_u = 5.69$ (with corresponding $x_{0u} = 3.44$) obtained by WPTS, showing excellent agreement with previous experimental and numerical results \cite{Tur-case-jet-exp-hussein1994velocity, Tur-case-jet-exp-panchapakesan1993turbulence, Tur-case-jet-DNS-sharan2021investigation}.

\begin{table}[h]
	\centering
	\begin{tabular}{ccc}
		\hline
		& $Re_j$ & $B_u$ \\
		\hline
		Exp Hussein & 100,000 & 5.80 \\
		\hline
		Exp Panchapakesan & 11,000 & 6.06 \\
		\hline
		DNS Sharan & 5,000 & 5.50 \\
		\hline
		Present WPTS & 5,000 & 5.69 \\
		\hline
	\end{tabular}
	\caption{The predicted $B_u$ by WPTS and comparison with previous studies.}
	\label{Tab-Bu}
\end{table}

\begin{figure}[htbp]
	\centering
	\subfigure{
		\includegraphics[height=6.0cm]{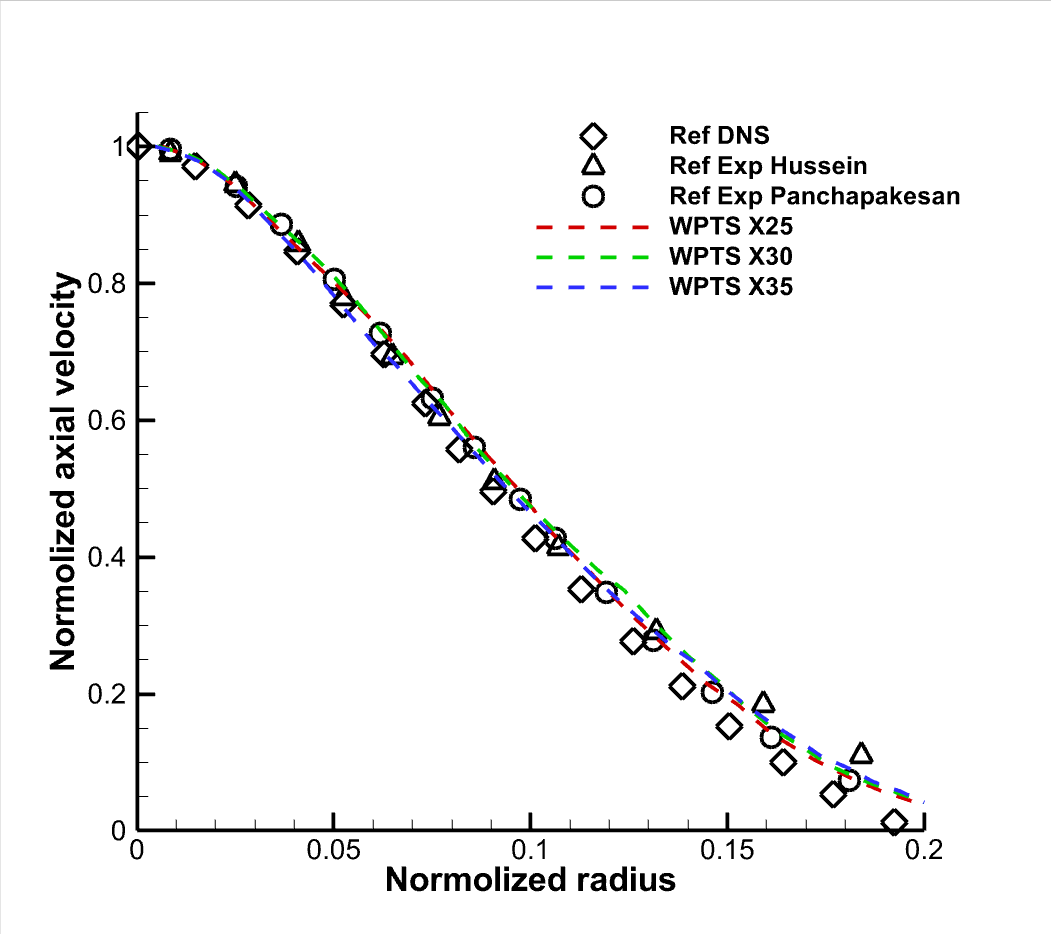}
	}
	\quad
	\subfigure{
		\includegraphics[height=6.0cm]{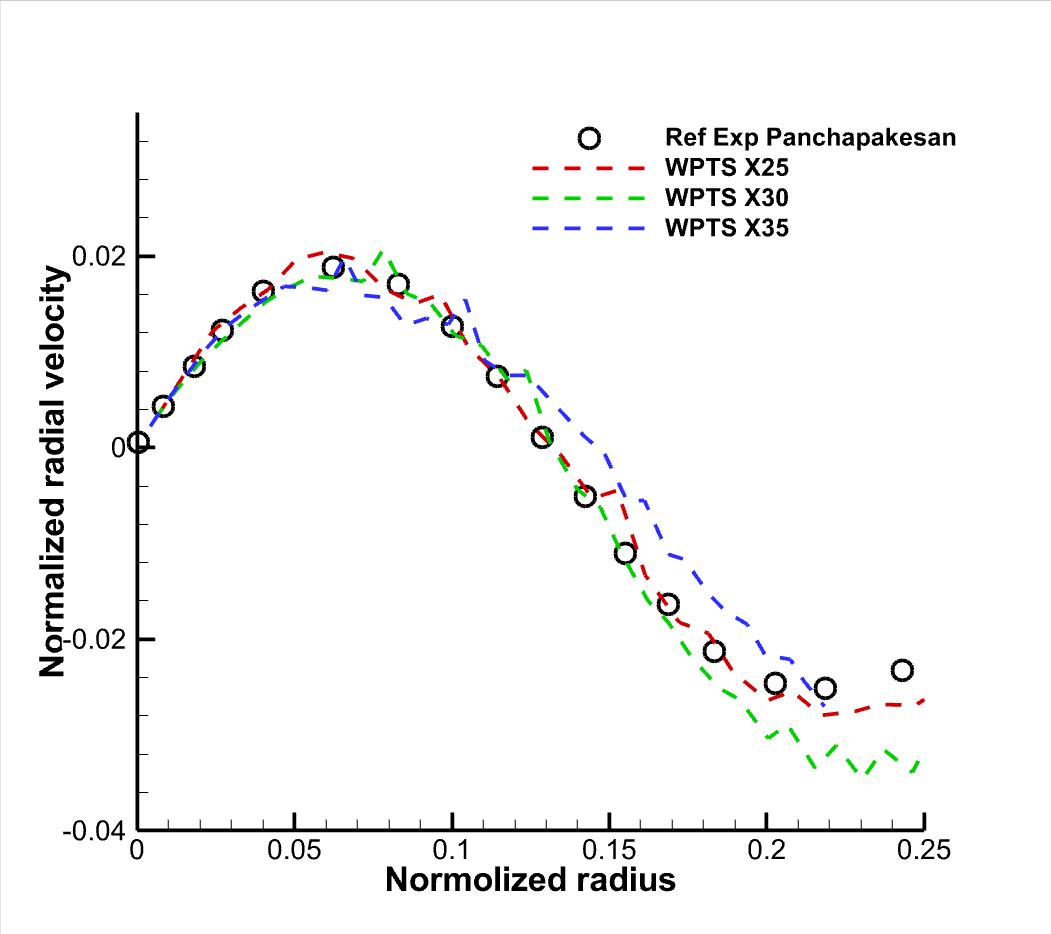}
	}
	\caption{The profiles of mean streamwise velocity and radial velocity.}
	\label{Fig-vel-mean}
\end{figure}

Next is the profile of mean velocity.
Both the mean streamwise velocity ($U$) and radial velocity ($U_r$) are analyzed, where the radial velocity can be obtained by transforming the $V$ and $W$ components into cylindrical coordinates. To obtain the mean value, the averaging is conducted in both temporal and azimuthal directions. Figure \ref{Fig-vel-mean} shows results at three streamwise locations, $x = 25, 30, 35$. It is worth noting that the mean velocity is normalized by the mean streamwise velocity at the centerline, $U_c\left(x\right)$, and the radius is normalized as $r_n = r/(x-x_{0u})$. The mean velocity profiles exhibit excellent collapse and agree well with reference results from experiments and DNS, particularly for the streamwise (or axial) component \cite{Tur-case-jet-exp-hussein1994velocity, Tur-case-jet-exp-panchapakesan1993turbulence, Tur-case-jet-DNS-sharan2021investigation}. Minor discrepancies in mean radial velocity for $r_n > 0.15$ are attributed to its significantly smaller magnitude, approximately two orders of magnitude less than the streamwise velocity, making these deviations acceptable given the small absolute values involved.

\begin{figure}[htbp]
	\centering
	\subfigure{
		\includegraphics[height=6.0cm]{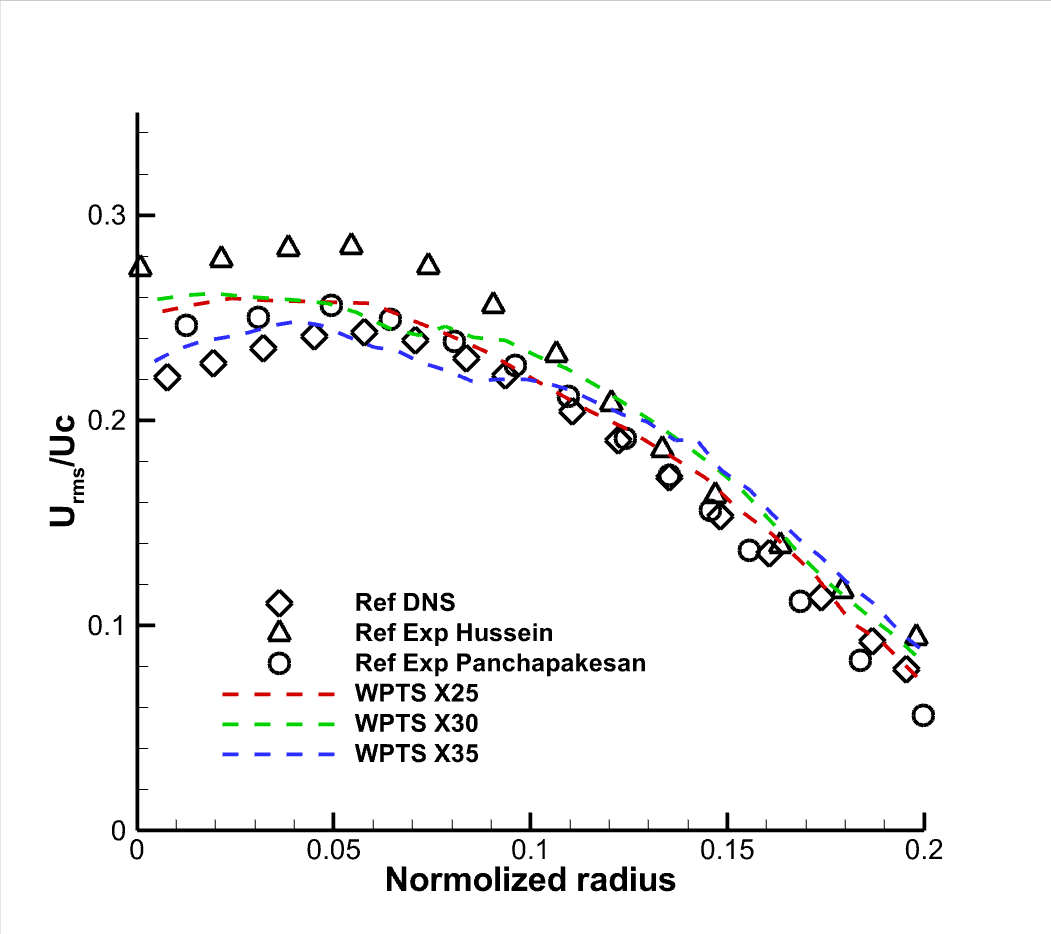}
	}
	\quad
	\subfigure{
		\includegraphics[height=6.0cm]{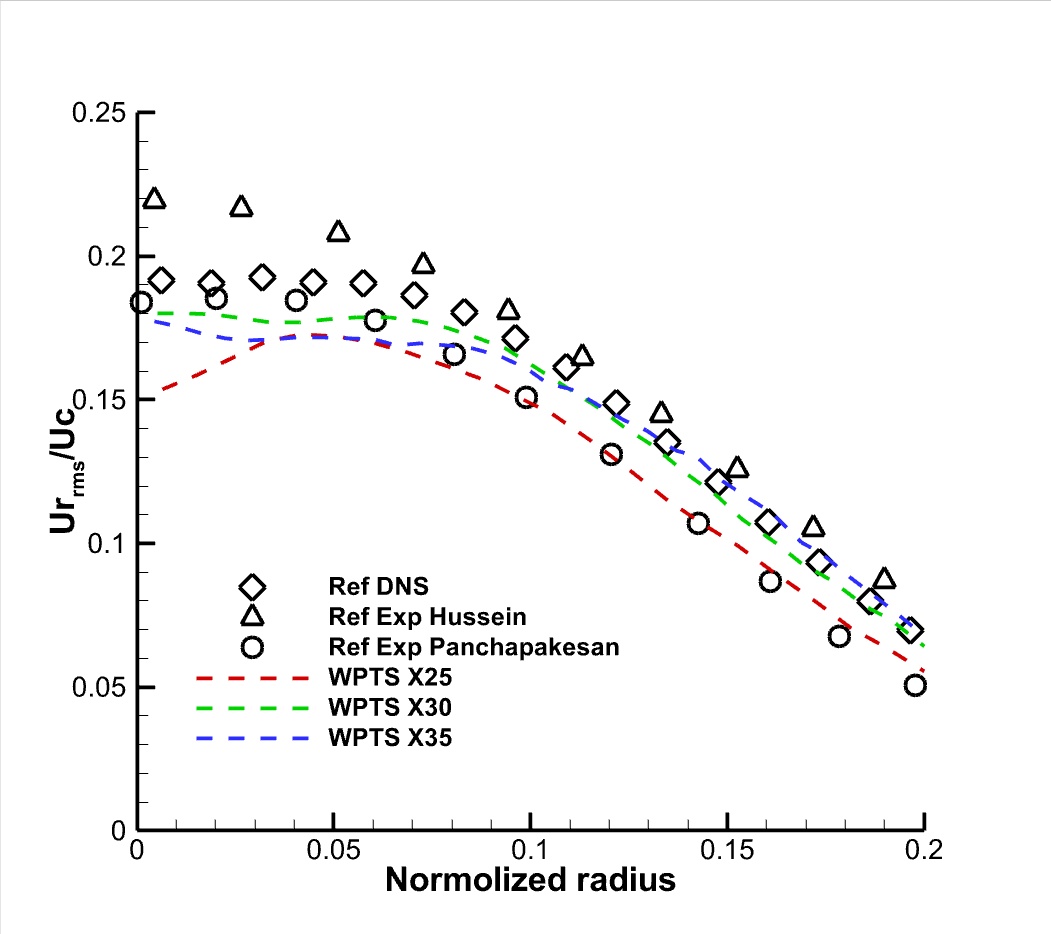}
	}
	\subfigure{
		\includegraphics[height=6.0cm]{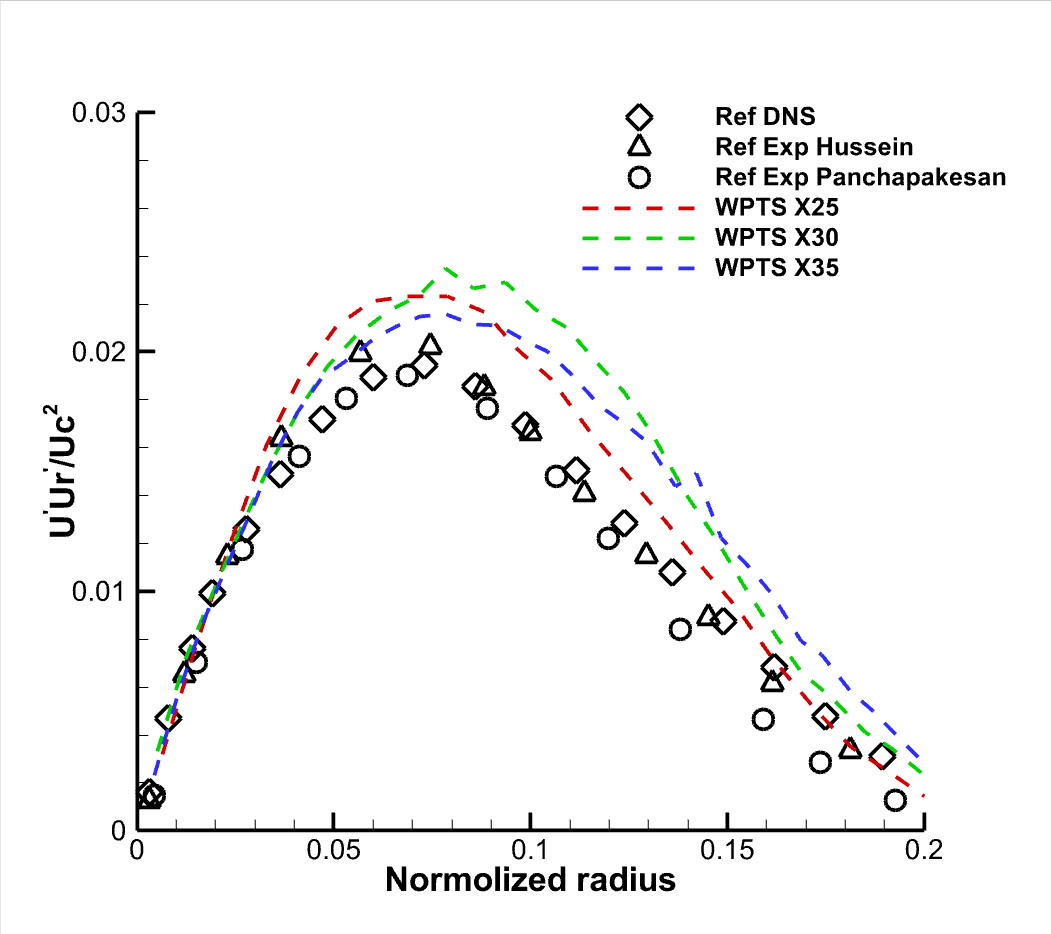}
	}
\caption{The profiles of Reynolds stress associated terms: (left) the r.m.s. of $U' U'$, (right) the r.m.s. of $U'_r U'_r$, (down) the cross-stress term of $U' U'_r$. $U$ and $U_r$ denote the velocity in the streamwise direction and radial direction, respectively.}
	\label{Fig-vel-rxx}
\end{figure}

Then it is about the Reynolds stress terms. Figure \ref{Fig-vel-rxx} presents the second-order turbulence statistics, specifically the Reynolds stress components, in the fully developed turbulent region. Results from the same three streamwise locations demonstrate satisfactory collapse and agreement with reference data \cite{Tur-case-jet-exp-hussein1994velocity, Tur-case-jet-exp-panchapakesan1993turbulence, Tur-case-jet-DNS-sharan2021investigation}, except slightly elevated cross-stress components, the $U^{'} U^{'}_r$ term. Overall, the WPTS method successfully predicts the anisotropic Reynolds stress distribution.
In summary, the linear decay rate, the mean velocity, and the distribution of Reynolds stress terms collectively demonstrate that WPTS is capable of accurately capturing the self-similarity characteristics of round jet flow, validating its reliability and accuracy.

\subsubsection{Flow features and wave-particle decomposition}
\begin{figure}[htbp]
	\centering
	\subfigure{
		\includegraphics[height=0.95cm]{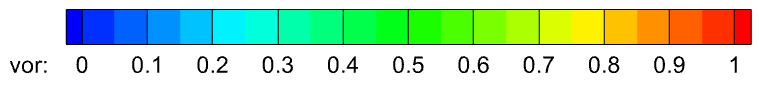}
	}
	\quad
	\subfigure{
		\includegraphics[height=7.0cm]{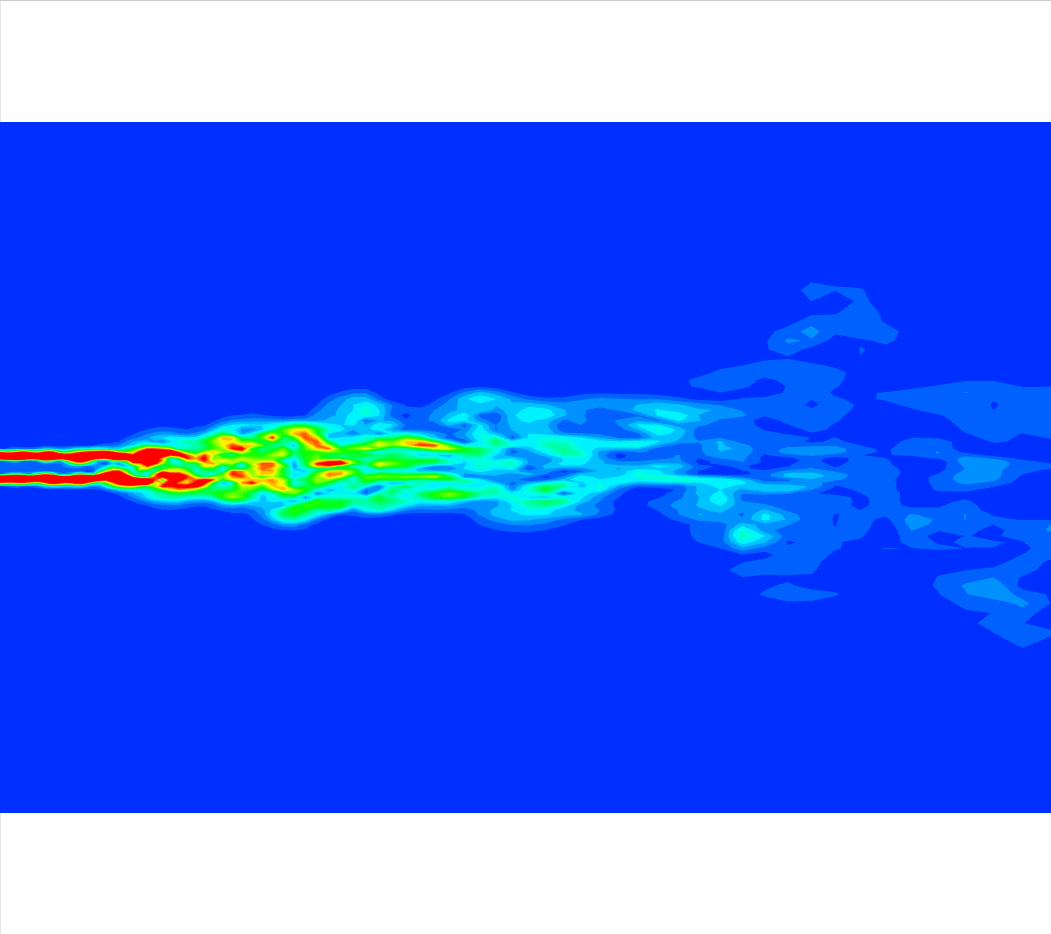}
	}
	\caption{The instantaneous snapshot of vorticity magnitude $|\vec{\omega}|$ at $t=20.0T_e$ for $xoy$ plane with $z=0$. The domain shown is $[0,45D]\times[-15D,15D]$. The $|\vec{\omega}|$ is defined by $|\vec{\omega}|=\sqrt{\omega_x^2 + \omega_y^2 + \omega_z^2}$.}
	\label{Fig-vor-z0}
\end{figure}

\begin{figure}[htbp]
	\centering
	\subfigure{
		\includegraphics[height=0.95cm]{figure/legend-vor-ho.png}
	}
	\\
	\subfigure{
		\includegraphics[height=4.0cm]{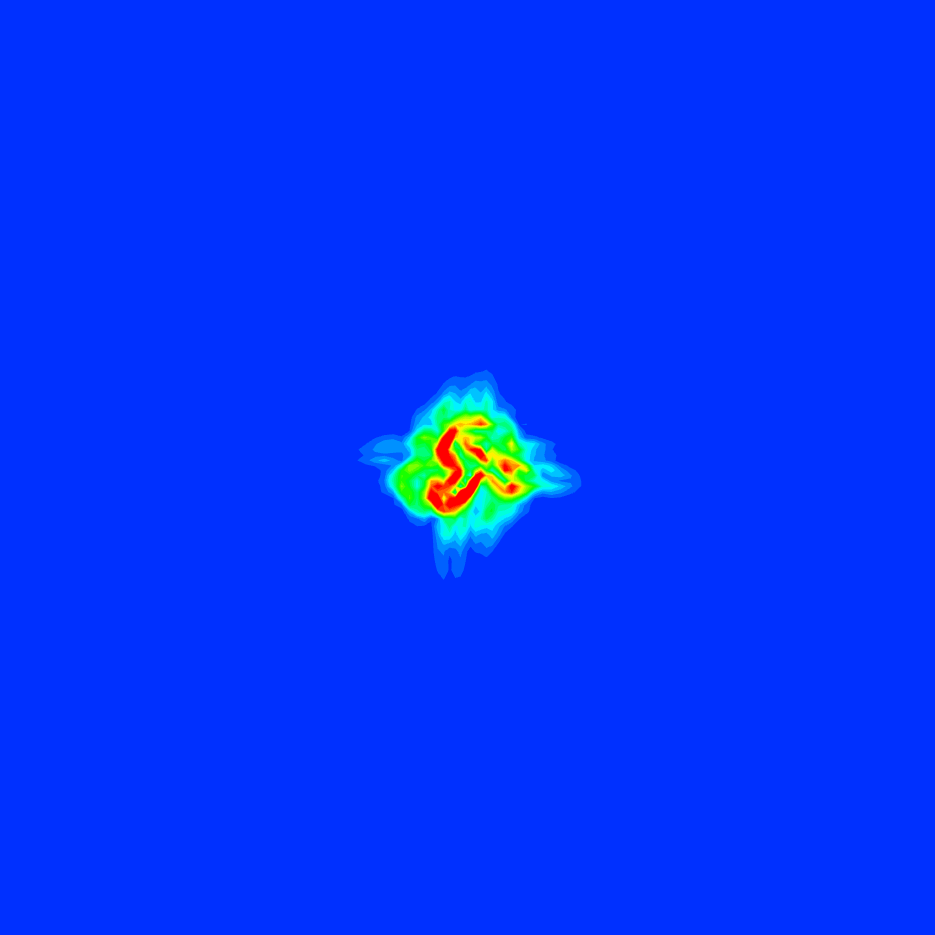}
	}
	\subfigure{
		\includegraphics[height=4.0cm]{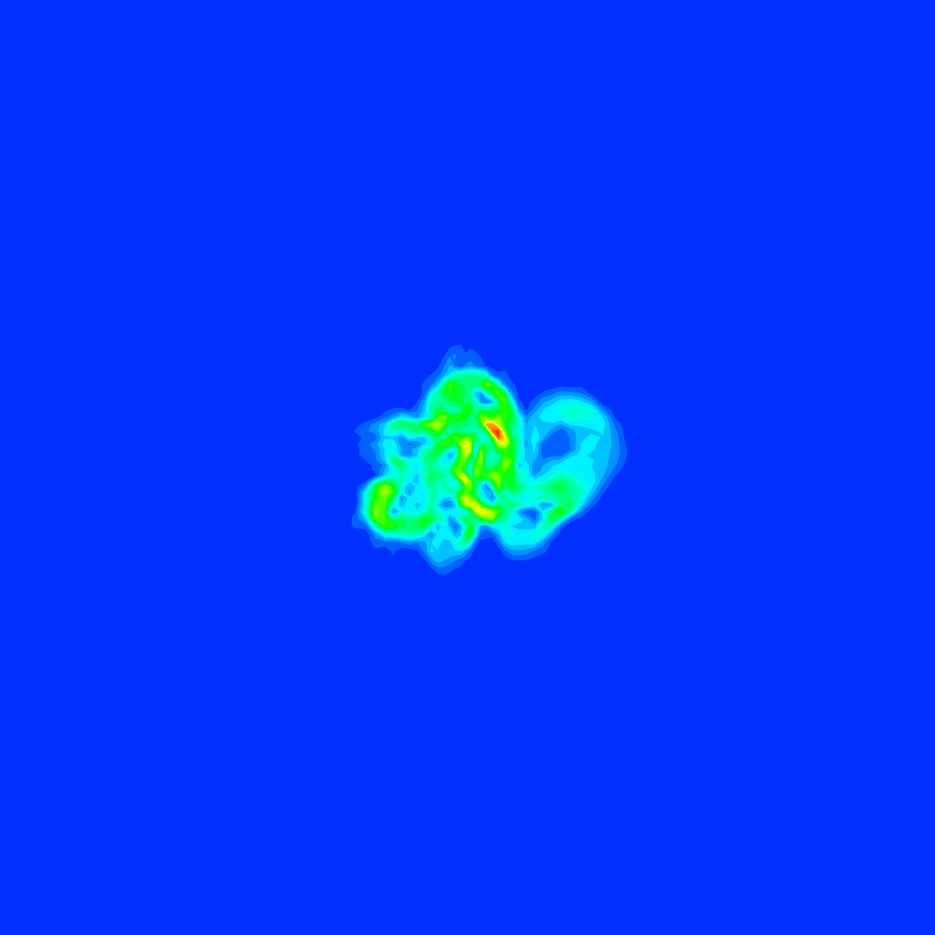}
	}
	\subfigure{
		\includegraphics[height=4.0cm]{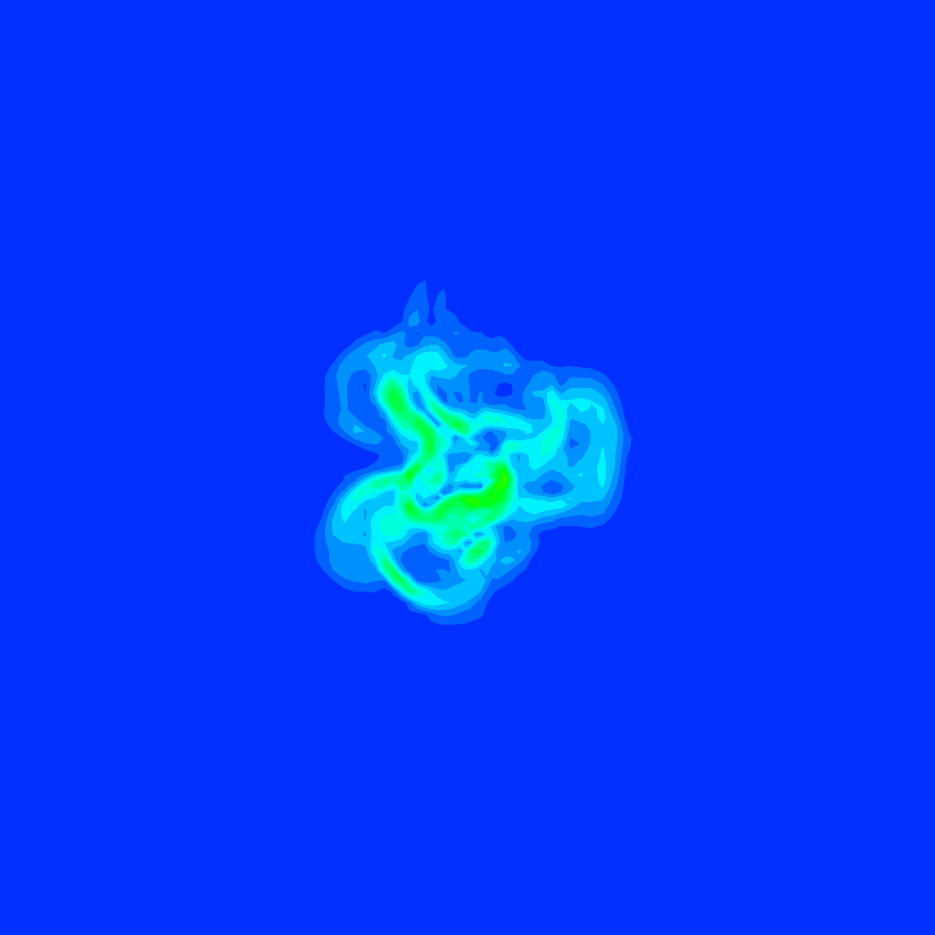}
	}
	\subfigure{
		\includegraphics[height=4.0cm]{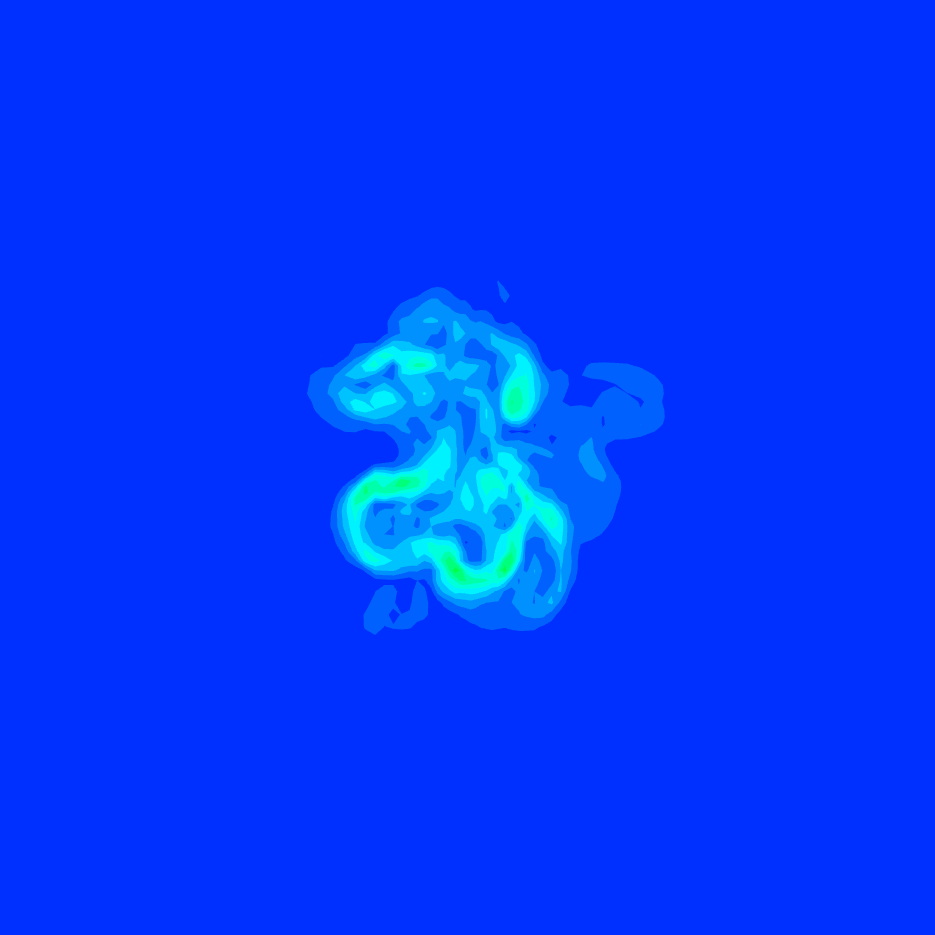}
	}
	\subfigure{
		\includegraphics[height=4.0cm]{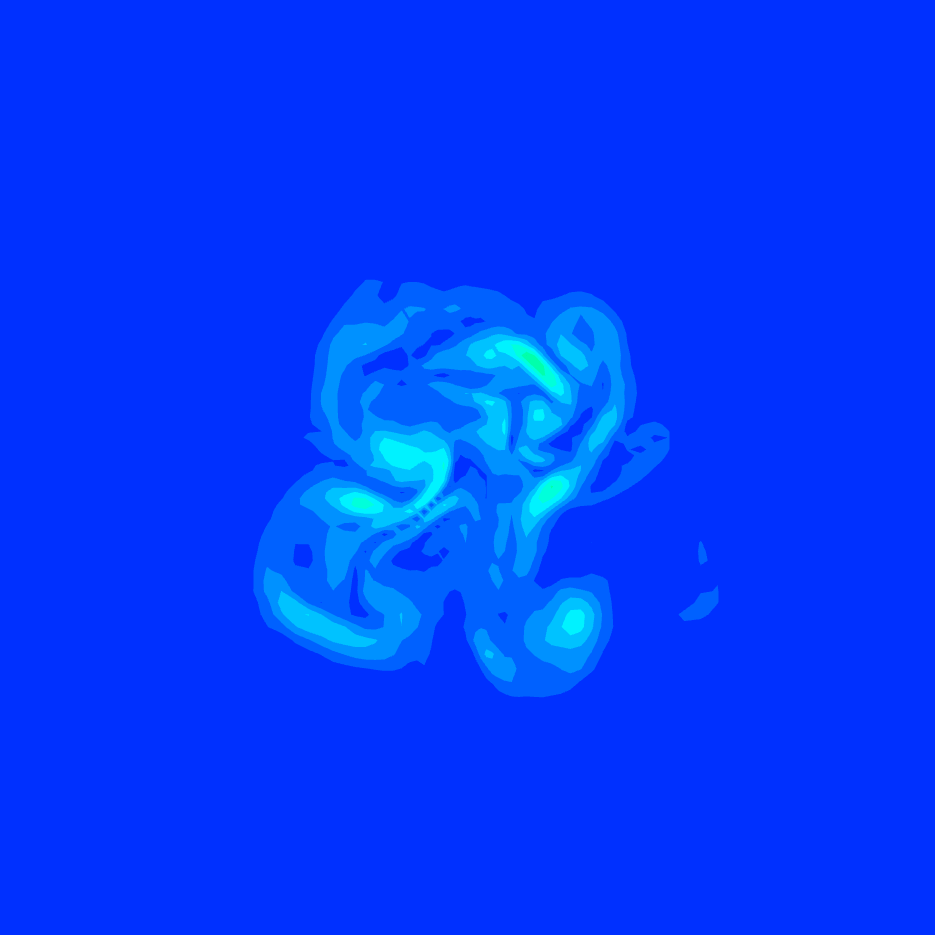}
	}
	\subfigure{
		\includegraphics[height=4.0cm]{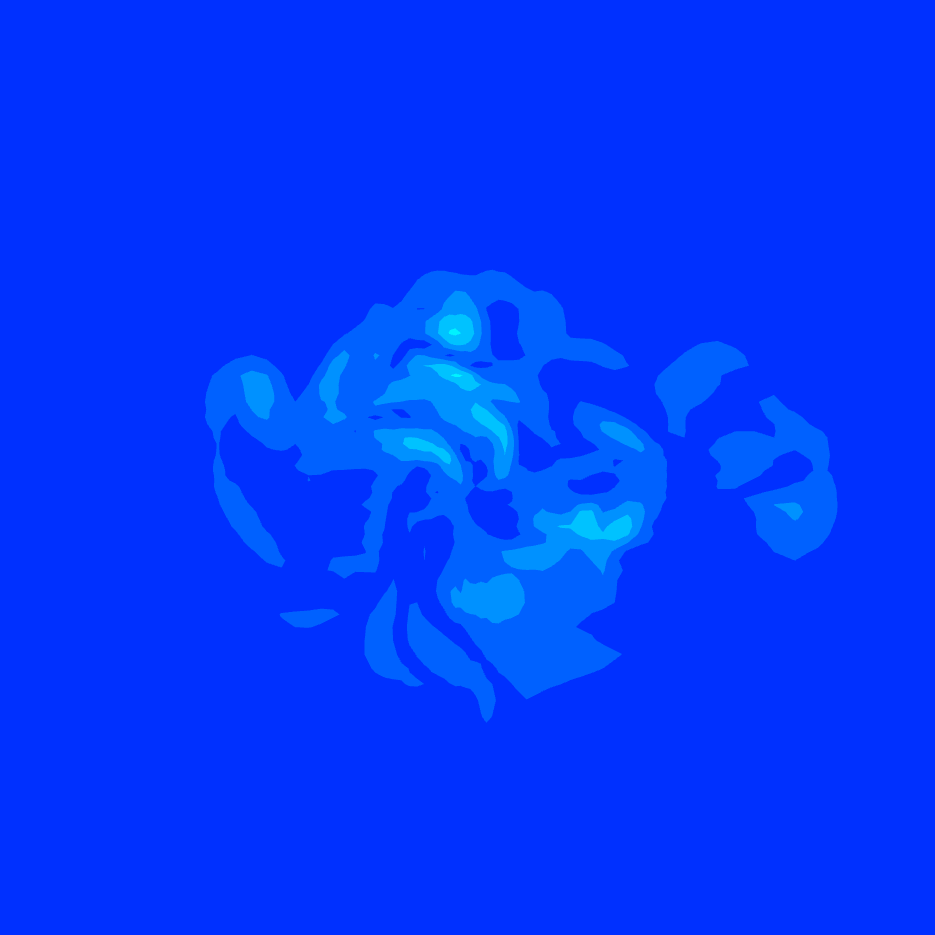}
	}
	\caption{The instantaneous snapshot of vorticity magnitude $|\vec{\omega}|$ at $t=20.0T_e$ for $yoz$ plane with $x=10D, 15D, 20D, 25D, 30D, 35D$, respectively. The domain shown is $[-10D,10D]\times[-10D,10D]$. The $|\vec{\omega}|$ is defined by $|\vec{\omega}|=\sqrt{\omega_x^2 + \omega_y^2 + \omega_z^2}$.}
	\label{Fig-vor-atx}
\end{figure}

Figure \ref{Fig-vor-z0} and Figure \ref{Fig-vor-atx} display the instantaneous distribution of vorticity in the $xoy$ and $yoz$ planes, respectively, at $t=20.0T_e$.
Consistent with previous studies, the vorticity magnitude peaks near the jet exit. Downstream development of the jet is characterized by a gradual attenuation of vorticity along the centerline (around $r=0$), coupled with spreading radially, as clearly demonstrated by the vorticity profiles at different streamwise locations in Figure \ref{Fig-vor-atx}. These results indicate the flow's evolving structure.

\begin{figure}[htbp]
	\centering
	\subfigure{
		\includegraphics[height=7.0cm]{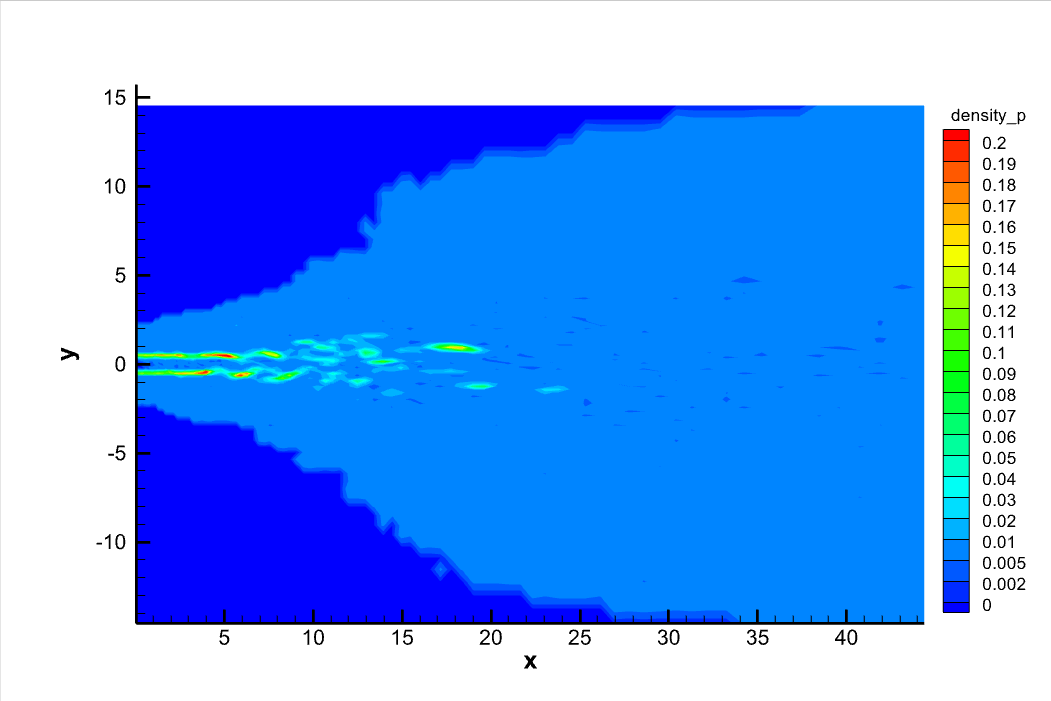}
	}
	\caption{The instantaneous snapshot of $\rho_p$, indicating the fluid represented by stochastic particles in WPTS, at $t=20.0T_e$ for $xoy$ plane with $z=0$. The domain shown is $[0,45D]\times[-15D,15D]$.}
	\label{Fig-rhop-z0}
\end{figure}

\begin{figure}[htbp]
	\centering
	\subfigure{
		\includegraphics[height=0.95cm]{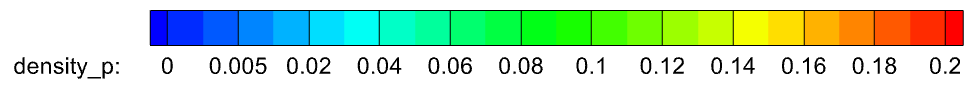}
	}
	\\
	\subfigure{
		\includegraphics[height=4.3cm]{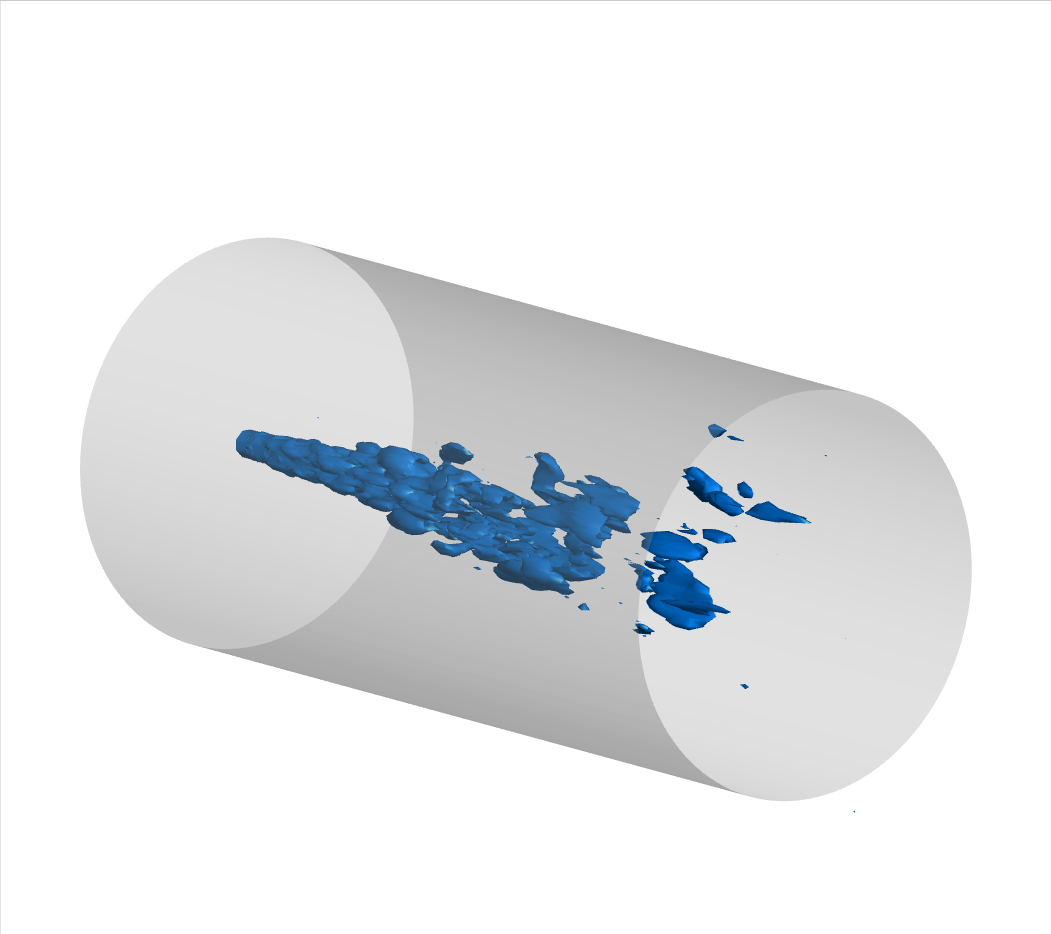}
	}
	\subfigure{
		\includegraphics[height=4.3cm]{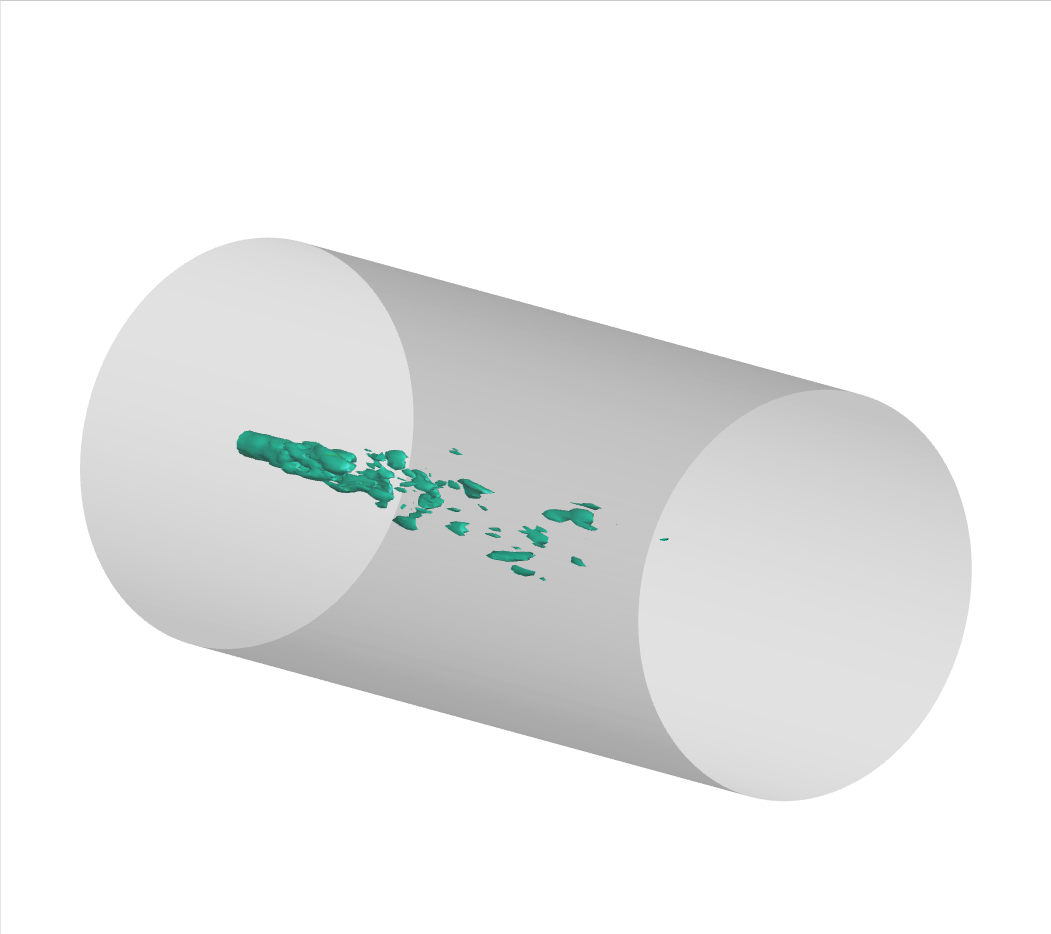}
	}
	\subfigure{
		\includegraphics[height=4.3cm]{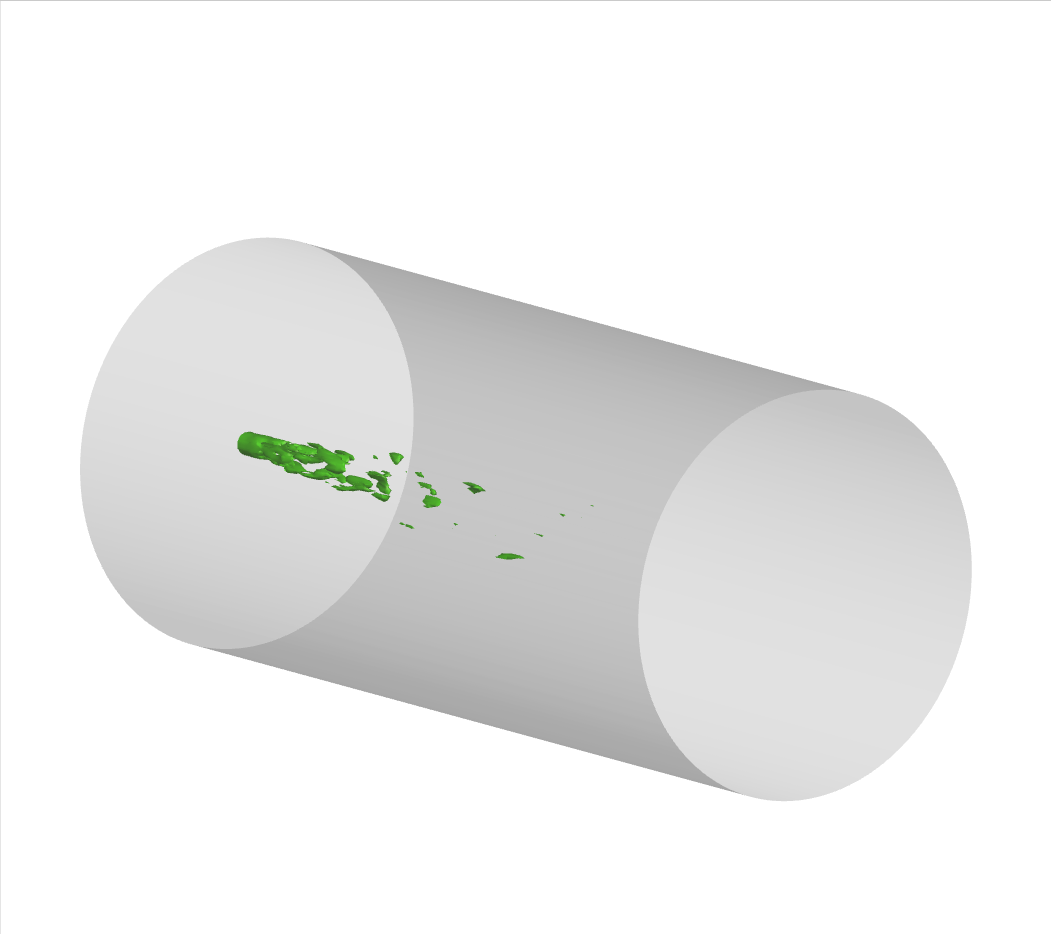}
	}
	\caption{The instantaneous isosurface of $\rho_p$, indicating the fluid represented by stochastic particles in WPTS, at $t=20.0T_e$. From left to right, the isosurfaces represent $\rho_p=0.01, 0.05, 0.10$, respectively. The gray translucent cylinder represents the region where $0 \le x \le 45D$ and $r \le 10D$ with $r=\sqrt{y^2+z^2}$.}
	\label{Fig-rhop-iso}
\end{figure}

As introduced previously, WPTS employs an adaptive wave-particle coupling algorithm mainly governed by the turbulent collision time $\tau_t$. Figure \ref{Fig-rhop-z0} illustrates the particle distribution in the $xoy$ plane with $z=0$. Generally, higher particle concentrations are found in regions of strong shear near the jet exit, where the strain rates resolved are high. In the downstream, as velocity gradients and turbulence intensity diminish gradually, the fluid evolved by stochastic particles in WPTS decreases accordingly. The similar trend can be further confirmed by the iso-surface of three-dimensional particle concentration shown in Figure \ref{Fig-rhop-iso}.
The adaptive feature mentioned above represents a key advantage of WPTS over the pure particle method (for turbulence simulation or modeling), as particles are selectively employed only in regions where grid resolution becomes inadequate for resolving small-scale turbulent structures. This approach maintains modeling flexibility while minimizing computational overhead, establishing WPTS as a powerful tool for turbulent flow simulations.
It is worth noting that the $E_p$ term in $\tau_n$ enhances the possibility of particle retention during evolution, which may explain the presence of particles in far-field regions.
Despite the low density of expressed by particles in some regions, their presence as a constituent of the fluid may have a non-negligible influence on fluid evolution, consequently affecting turbulent statistics variables.

\subsubsection{Sensitivity to frequency ratio in inlet perturbation}
\begin{figure}[htbp]
	\centering
	\subfigure{
		\includegraphics[height=6.5cm]{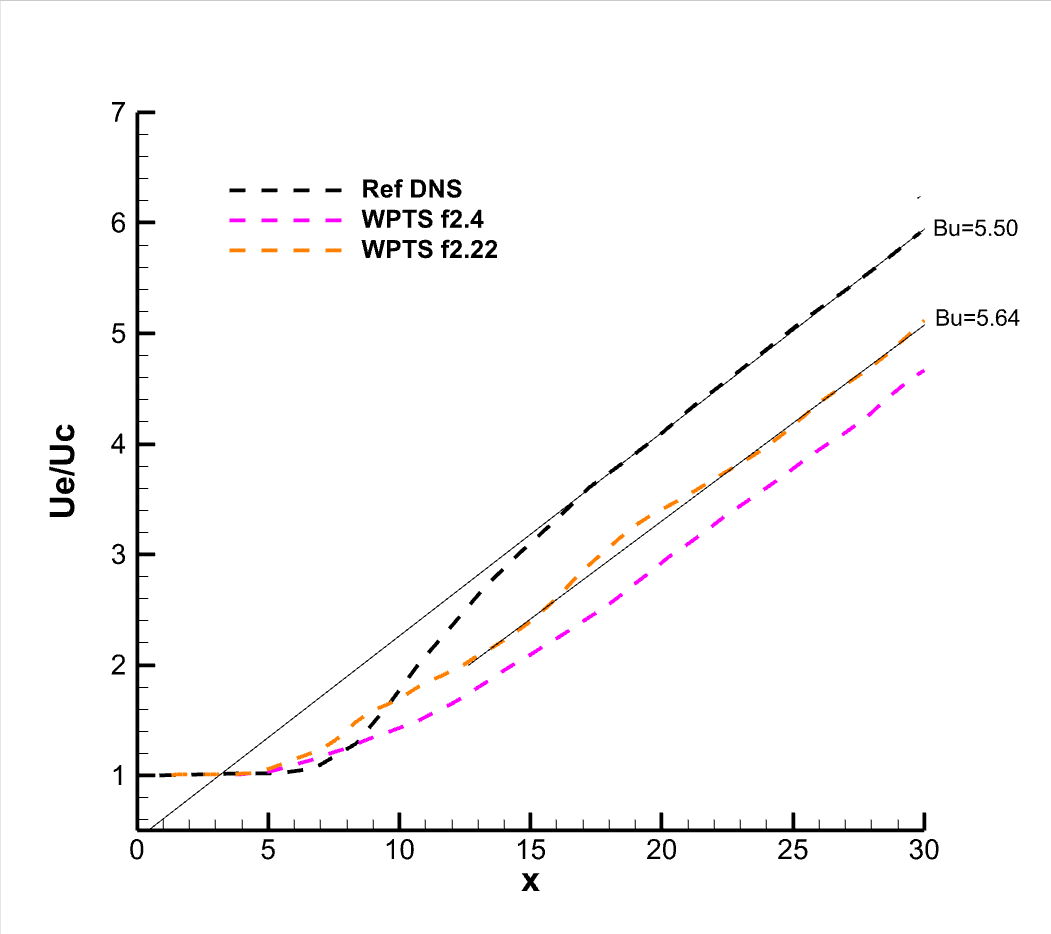}
	}
	\caption{The inverse of averaged centerline mean velocity in the case of $f = 2.22$, and the reference is from \cite{Tur-case-jet-DNS-sharan2021investigation}.}
	\label{Fig-f2d22-Bu}
\end{figure}

\begin{figure}[htbp]
	\centering
	\subfigure{
		\includegraphics[height=6.0cm]{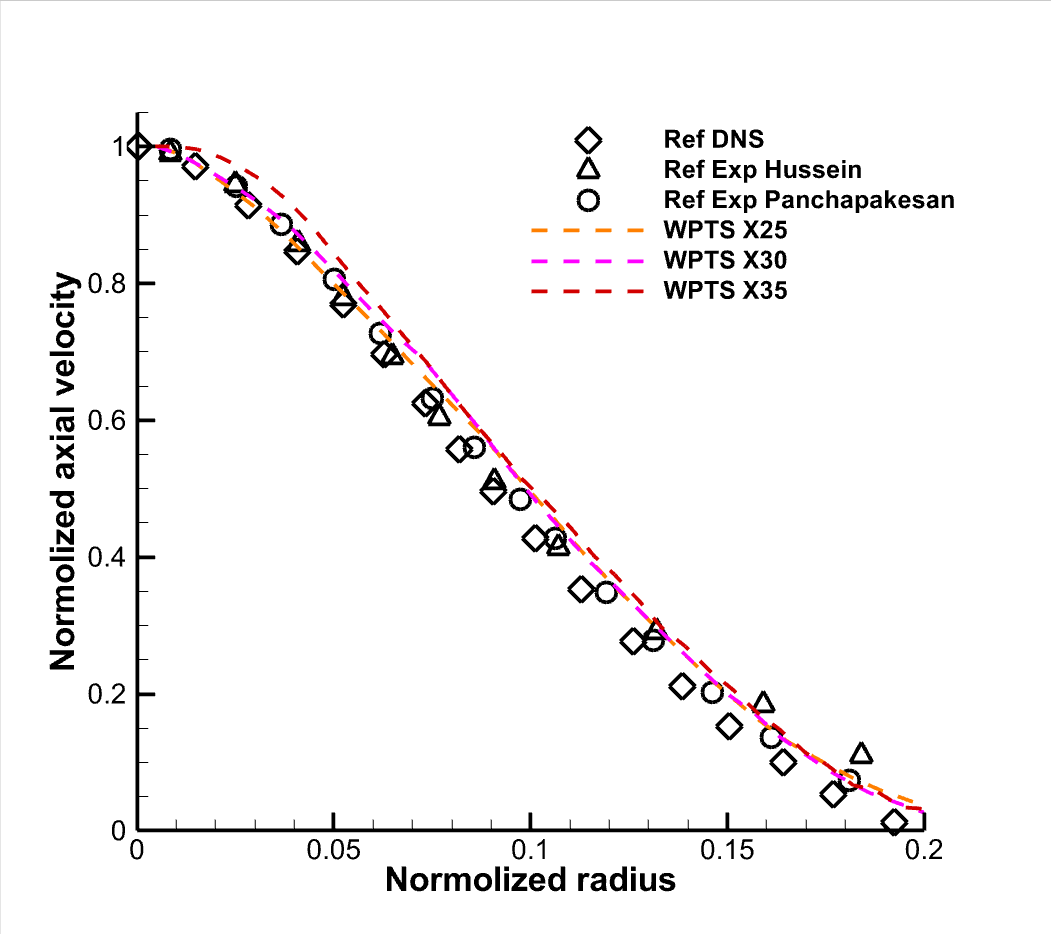}
	}
	\subfigure{
		\includegraphics[height=6.0cm]{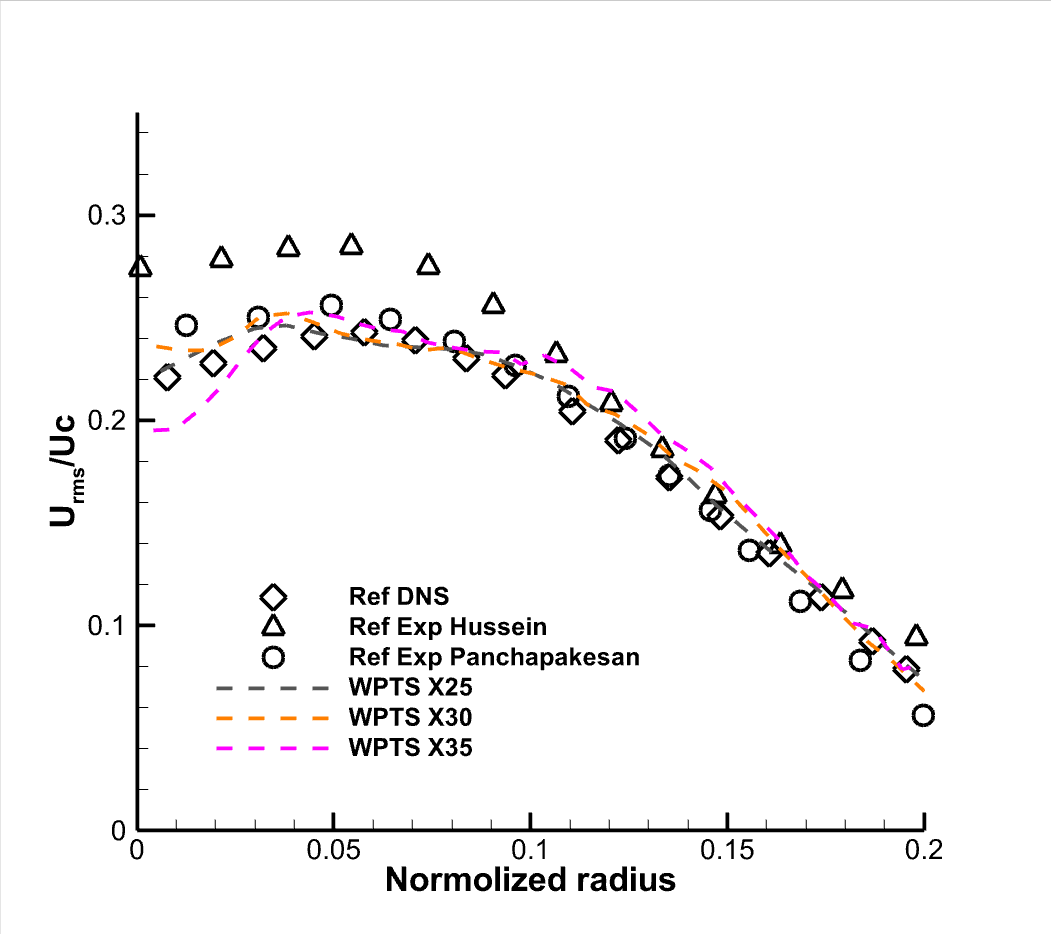}
	}
	\subfigure{
		\includegraphics[height=6.0cm]{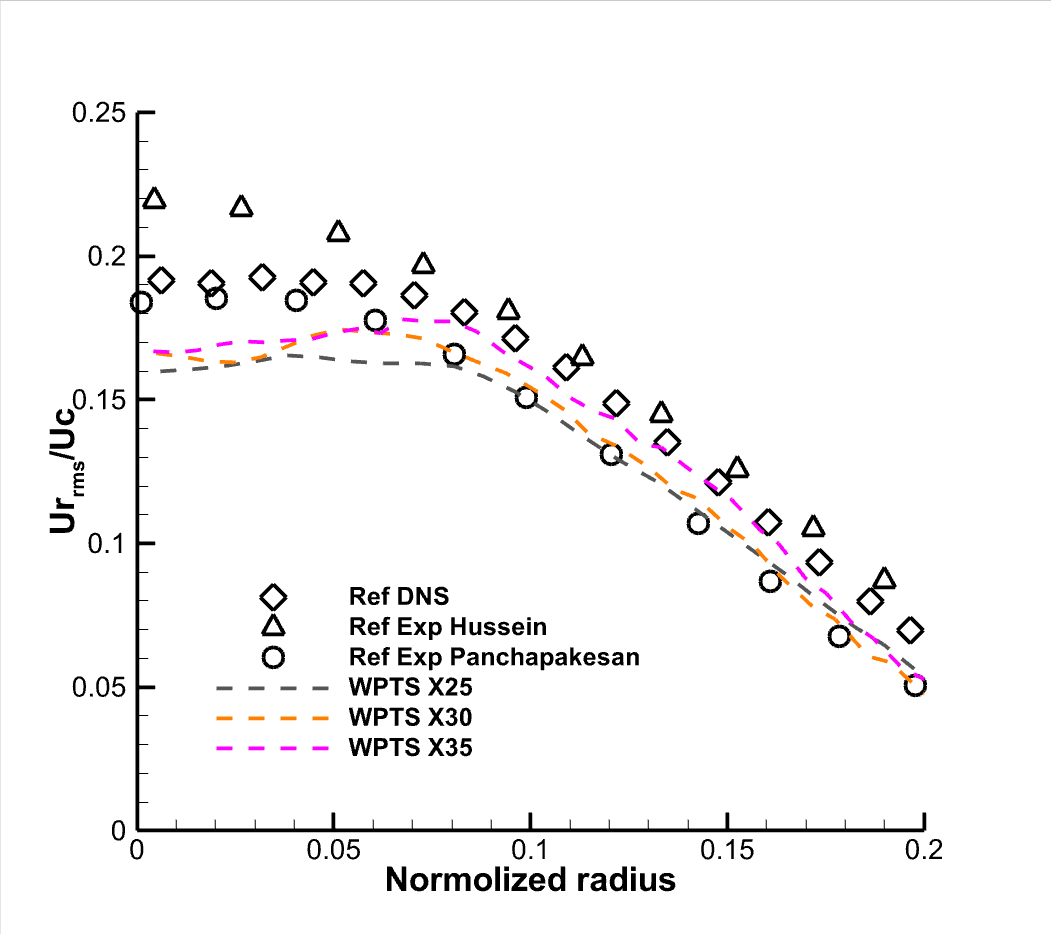}
	}
	\subfigure{
		\includegraphics[height=6.0cm]{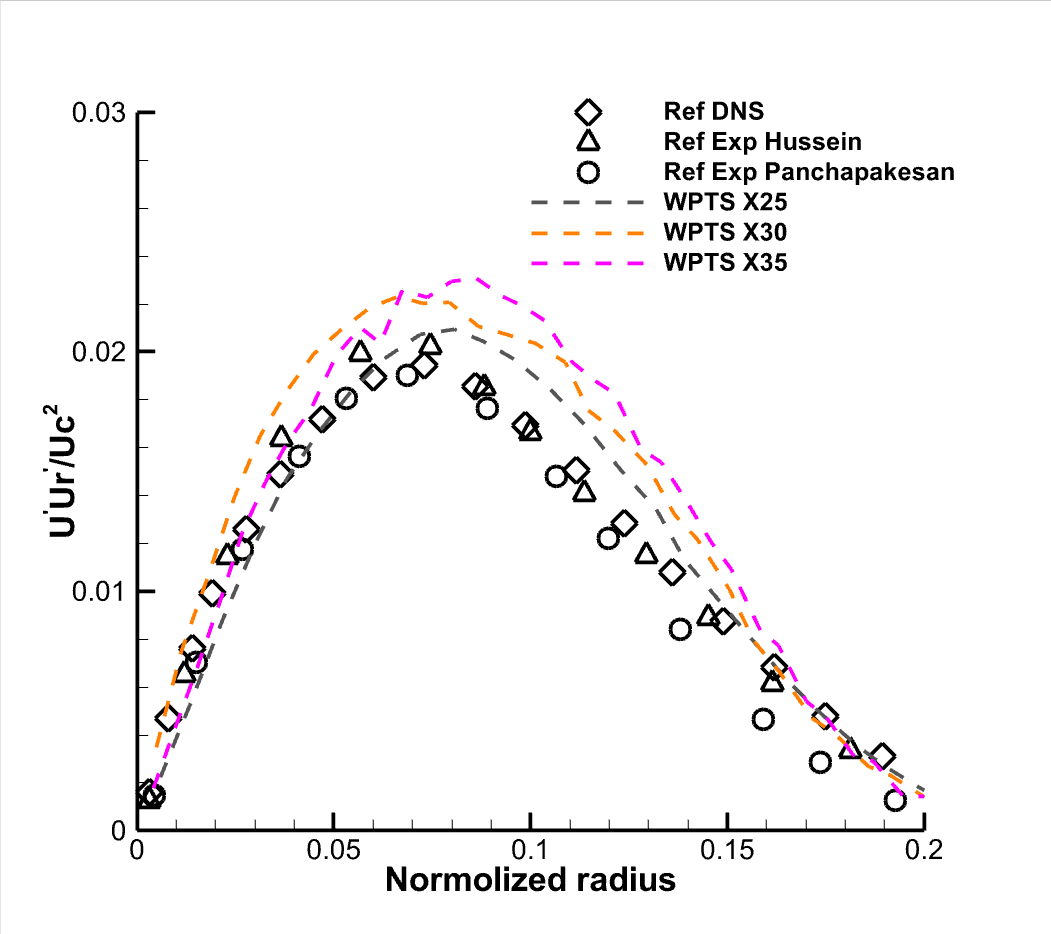}
	}
	\caption{The profiles of mean velocity and Reynolds stress associated terms in the case of $f = 2.22$: the mean streamwise velocity $U$, the r.m.s. of $U' U'$, the r.m.s. of $U'_r U'_r$, the cross-stress term of $U' U'_r$. $U$ and $U_r$ denote the velocity in the streamwise direction and radial direction, respectively.}
	\label{Fig-f2d22-urxx}
\end{figure}

Additional simulations are further performed to examine the influence of the frequency ratio $f$ in the inlet perturbation Eq.\eqref{eq-uInletmode} on WPTS. Following the study in \cite{Tur-case-jet-inflow-gohil2015simulation}, we tested an alternative inlet condition with $f = 2.22$ while other coefficients keep the same, namely, $A_n=A_h=0.05$, and $St_D = 0.5$. Figure \ref{Fig-f2d22-Bu} demonstrates that the trend of linear decay is also well-predicted, which indicates $B_u = 5.64$, consistent with previous results. Different from the result with $f = 2.40$, the curve of $U_e/U_c$ - $x$ now shows a slight upward shift after around $x = 10D$, resulting in the obtained $x_{0u} = 1.34$. It suggests that the frequency ratio in inlet perturbation primarily affects the transition region from jet exit to the fully-developed flow, without altering the characteristics in the fully-developed turbulence, i.e., the self-similarity feature, which can be further confirmed in Figure \ref{Fig-f2d22-urxx}, presenting the mean velocity and Reynolds stress statistic terms. Overall, the results show minimal sensitivity of predicted flow variables by the WPTS method to the given variation in inlet frequency.

\section{Conclusion}

In this paper, we have applied the recently proposed WPTS method to spatially developing jet flows. Unlike DNS, which requires highly refined grids to resolve Kolmogorov-scale structures, WPTS employs a multiscale approach that operates effectively on coarse grids. Departing from traditional LES and RANS modeling paradigms, WPTS decomposes the flow field into coupled wave and particle components. The wave component captures grid-resolved large-scale structures, while the particle component models sub-grid flow information through stochastic transport, with both components evolving through bidirectional coupling.

A key innovation of WPTS lies in its direct modeling approach: particle behaviors—including generation, transport, and dissipation—intrinsically represent the physical processes of unresolved turbulent kinetic energy production, propagation, and dissipation. Notably, WPTS exhibits domain-adaptive behavior through appropriate modeling of the turbulent relaxation time $\tau_t$. In regions where the grid adequately resolves flow structures, WPTS automatically reduces to the gas-kinetic scheme (GKS), recovering the Navier-Stokes solution for laminar or well-resolved flow regions.

We validated the method using a jet flow at Reynolds number $5000$ and Mach number $0.6$. Despite employing only $2\%$ of the grid cells required for DNS, WPTS accurately captures the characteristic linear decay of centerline velocity in the streamwise direction. Key turbulence statistics at various axial locations—including mean velocity profiles and Reynolds stress components—demonstrate excellent agreement with both DNS data and experimental measurements. These results confirm that WPTS provides accurate and reliable predictions of jet flow behavior while offering substantial computational efficiency. The successful application to this canonical turbulent flow configuration establishes WPTS as a promising tool for simulating more complex engineering flows where computational resources are constrained.

\section{Acknowledgements}
The current research is supported by National Key R\&D Program of China (Grant Nos. 2022YFA1004500), National Science Foundation of China (12172316, 92371107), and Hong Kong research grant council (16301222, 16208324).

\bibliographystyle{plain}%
\bibliography{reference}
\end{document}